\documentclass[iop]{emulateapj}
\usepackage{natbib}
\usepackage{graphicx,color,rotating}
\usepackage{footnote}
\usepackage{ulem} 
\usepackage{graphicx,amssymb,amsmath,amsfonts,times,hyperref}

\shorttitle{Diffuse $\gamma$-ray emission from misaligned AGN}
\shortauthors{Di Mauro et al.}

\begin{document}
   \title{Diffuse $\gamma$-ray emission from misaligned active galactic nuclei}
\author{M. Di Mauro}
\affil{Physics Department, Torino University, and  Istituto Nazionale di Fisica Nucleare, Sezione di Torino, via Giuria 1, 10125 Torino, Italy}
   \author{F. Calore}
   \affil{II. Institute for Theoretical Physics, University of Hamburg, Luruper
Chaussee 149, 22761 Hamburg, Germany}
   \author{ F. Donato\altaffilmark{1}}
   \affil{Physics Department, Torino University, and  Istituto Nazionale di Fisica Nucleare, Sezione di Torino, via Giuria 1, 10125 Torino, Italy}
\author{M. Ajello}
\affil{Space Sciences Laboratory, University of California,
Berkeley, CA, 94720}
\author{L. Latronico}
\affil{Istituto Nazionale di Fisica Nucleare, Sezione di Torino, via Giuria 1, 10125 Torino, Italy}	  

\altaffiltext{1}{corresponding author: donato@to.infn.it}

  \begin{abstract}
  Active galactic nuclei (AGN) with 
 jets seen at small viewing angles are the most luminous and abundant objects in the $\gamma$-ray sky.
 AGN with jets misaligned along the line-of-sight appear fainter in the sky, but are more numerous than the brighter blazars. 
We calculate the diffuse $\gamma$-ray emission due to the population of misaligned AGN (MAGN) unresolved 
by the Large Area Telescope (LAT) on the {\it Fermi} Gamma-ray Space Telescope ({\it Fermi}). A  correlation between the $\gamma$-ray luminosity 
and the radio-core luminosity is established and  demonstrated  to be physical by statistical tests, as well as 
compatible with upper limits based on {\it Fermi}-LAT data for a large sample of radio-loud MAGN. We constrain the derived
$\gamma$-ray  luminosity function by means of the source count distribution of the radio galaxies (RGs)
 detected by the {\it Fermi}-LAT. 
We finally calculate the diffuse  $\gamma$-ray flux due to the whole MAGN population. 
Our results demonstrate that the MAGN can contribute from 10\% up to nearly the entire measured Isotropic Gamma-Ray Background (IGRB). 
We evaluate a theoretical uncertainty on the flux of almost an order of magnitude.
  \end{abstract}
  

   \maketitle


\section{Introduction}

The {\it Fermi}-LAT has measured the Isotropic Gamma-Ray Background (IGRB) with very good 
accuracy from 200 MeV to 100 GeV  \citep{IDGRB}.
 Increased statistics from additional LAT data are expected to allow the IGRB to be measured over an
even broader energy range. 
The nature of the IGRB is still an open problem in astrophysics. 
Blazars  and star-forming galaxies contribute 20\%-30\% of the IGRB \citep{2010ApJ...720..435A,2012ApJ...755..164A},
 and with a compatible spectral slope. Blazars are Active Galactic Nuclei (AGN) 
whose jets are oriented along the lines-of-sight (l.o.s.). 
Their luminosity is quite high, due to Doppler boosting effects. For the same reasons, 
 AGN with axes misaligned with respect to the line-of-sight (hereafter MAGN) 
 have weaker luminosities but are expected to be more numerous by a factor ~2$\Gamma^2_L$ (where $\Gamma_L$ is the Lorentz factor)
 \citep{urry1995}.
 About  $10\%$ of the observed AGN are radio-loud. According to the unified model, 
AGN are classified as a function of their jet orientation with respect to the l.o.s..
A jet misalignment of about 14$^\circ$ indicates the separation between 
blazars and non-blazar, $i.e.$ misaligned, objects \citep{urry1995}. In the unified model, radio galaxies (RGs)
are those objects which, on average, have their jets pointing
at $>44^\circ$ from our l.o.s., while MAGN below this angle and above about 14$^\circ$  
are generally identified with radio quasars \citep{1989ApJ...336..606B}.
RGs are classified in turn into two categories based upon their radio morphology  \citep{fanaroff1974}. 
The first class of objects, named Fanaroff-Riley Type I (FRI), 
is preferentially found in rich clusters and hosted by weak-lined galaxies, and has a low luminosity radio 
emission (which peaks near the center of the AGN and shows two-sided jets dominated emission).
 Fanaroff-Riley Type II (FRII) galaxies present a high luminosity radio emission dominated by the lobes,  while jets and core, when detected, are faint. 
The hot spots, generally not present in FRIs, are usually detected at the ends of the lobes.
The threshold in luminosity for discriminating between FRI and FRII is
about $10^{25}$ W Hz$^{-1}$ sr$^{-1}$ at 178 MHz \citep{fanaroff1974}.
FRIs and FRIIs are considered the parent populations of BL Lacertae objects (BL Lacs) and flat spectrum radio quasars (FSRQs), respectively.
 
A recent analysis finds that the FIRST radio sources undetected by the {\it Fermi}-LAT 
may explain about one half of the IGRB \citep{2013ApJ...769..153Z}.
The contribution of unresolved blazars to the IGRB 
has been shown to be non negligible \citep{stecker1993,1993MNRAS.260L..21P,salamon1994,2011PhRvD..84j3007A,2010ApJ...720..435A} 
and able to explain at least 10\% of the measured  IGRB at high latitudes \cite{IDGRB}.  
Given the large numbers of known radio-loud MAGN, and in  analogy with 
blazars,we estimate in this work how the faint but numerous 
unresolved MAGN population may contribute to the IGRB at a non-negligible level \citep{stawarz,2009RAA.....9.1205B,inoue}.
Using the first year {\it Fermi}-LAT catalog, \cite{inoue} evaluated the contribution of 
misaligned AGNs to be ranging between 10 and 63\% of the IGRB.
 We investigate the absolute
level of the  MAGN $\gamma$-ray flux and quantify the possible uncertainties of our prediction. 
The main original points of our analysis include 
i) the derivation of 
a $\gamma$-ray - radio core luminosity correlation for the MAGN detected by {\it Fermi}-LAT; 
ii) a check of this correlation against upper limits from tens of radio loud MAGN undetected in $\gamma$-rays; 
iii) statistical tests that verify that the evaluation of the  radio core - $\gamma$-ray luminosity correlation is not spurious; 
iv) the computation of the $\gamma$-ray luminosity function from the core radio one; 
v) evaluation of  the uncertainties affecting $\gamma$-ray flux predicted from the unresolved MAGN population.

The paper is organized as follows: in Sect.  \ref{sec:correlation} we derive the correlation between radio core ($L_{r, {\rm core}}$) 
and $\gamma$-ray luminosities ($L_{\gamma} $) from a sample of 12 MAGN detected by {\it Fermi}-LAT. 
The robustness of the $L_{\gamma} - L_{r, {\rm core}}$ correlation is tested in Sect.  \ref{sec:UL}
 by computing the 95$\%$ confidence level (C.L.) upper limits on the
 $\gamma$-ray flux of a sample of radio-loud MAGN with 4-years of {\it Fermi} data. 
 In Sect. \ref{sec:testcorr} two statistical tests are performed on the core radio and $\gamma$-ray data in order to exclude spurious effects in 
 the correlation between luminosities. 
 By assuming the relation found between $L_{\gamma} $ and $L_{r, {\rm core}}$, in Sect.  \ref{sec:GLF} we model the $\gamma$-ray 
 luminosity function (GLF) from the radio luminosity function (RLF).
We discuss the consistency of the models  in Sect.  \ref{sec:Ncount}, where we compare our predictions 
of the source count distribution and compare them 
to the {\it Fermi}-LAT data. Our findings for the contribution of an unresolved population of MAGN to the IGRB are presented in Sect.  \ref{sec:flux}, 
together with the evaluation of the relevant uncertainties. Finally, we draw our conclusions in Sect.  \ref{sec:conclusions}.
\\
Throughout the paper we adopt a standard $\Lambda$CDM cosmology with
parameters: $H_{0} = 70$ km s$^{-1}$ Mpc$^{-1}$, $\Omega_{M}$ = 0.27, $\Omega_{\Lambda}$ = 0.73.

\section{The correlation between $\gamma$-ray and radio luminosity}
\label{sec:correlation} 
The calculation of the diffuse emission from unresolved ($i.e.$ not detected by the {\it Fermi}-LAT) 
MAGN relies on the $\gamma$-ray luminosity function for that specific population. 
The physical processes underlying the emission of $\gamma$ rays in RGs are not 
firmly established. 
However, in analogy with blazars - being the same objects with off-line axes -  it is commonly assumed that MAGN 
experience their same emission processes \citep{2012IJMPS...8...25G}. 
It is believed that the bulk of the radiation is generated via synchrotron self-Compton (SSC) scatterings, 
where the seed photons are provided by synchrotron emission by the same electron population \citep{1992ApJ...397L...5M}. 
It is not excluded that an external inverse Compton (EC) scattering occurs off photons external to the jet \citep{1993ApJ...416..458D}. 
Dedicated studies of M87 \citep{Abdo:2009ta}, Cen A \citep{Falcone:2010fk} and NGC 1275 \citep{2009ApJ...699...31A} 
show that the SSC process successfully fits the
observed emission on a wide photon energy range, even if other mechanisms have been explored \citep{2012PhLB..707..255K}.

The contribution of kpc-scale jets and radio lobes to the IGRB is less than 10\,\%,
 as shown by   \citep{stawarz,2011ApJ...729L..12M}.
Ultra-relativistic electrons in the lobes emit synchrotron radiation in the radio band and are able to up-scatter low energy photons via IC scattering to high 
energies, provided a high enough electron density is available. 
The dominant contribution is expected to be from CMB photons. The IC/CMB scattered emission in
 the lobes of distant galaxies is generally well observed in the X-ray band. Extended $\gamma$-ray emission
 spatially coincident with  radio lobes has been detected from Centaurus A \citep{centaurus_a}. 
Such emission, if interpreted in terms of IC scattering of electrons with ambient photons, 
 requires high-energy electrons in the lobes, but it is  unclear how common this is in other RGs.
In what follows, we assume the $\gamma$-ray radiation originates in the central region of the source, as is predicted from both SSC and EC scenarios. 

The FRI and FRII galaxies show a strong emission in a wide radio band, 
spanning from hundreds of MHz up to tens of GHz. 
These photons are ascribed to the synchrotron emission of highly relativistic electrons moving 
 in the entire region of the source. 
The total radio flux has been measured for hundreds of FRI and FRII galaxies. 
For a number of these galaxies the emission from the central
unresolved region of an arcsecond scale, often referred to as the core, is detected as well.
In the first and the second catalogs of LAT AGN sources \citep{misaligned,secondcatalogAGN}
{\it Fermi}-LAT has reported the detection of 15 MAGN, which can be classified into 10 FRI and 5 FRII 
 galaxies (although with some caveats, see below).
 \cite{misaligned} report on the observation of 3C 78, PKS 0625-35, 3C 207, 3C 274, Centaurus A, NGC 6251, 
3C 380, 3C 120,  3C 111, 3C 84, PKS 0943-76, while Centaurus B, Fornax A and IC 310 have been reported in the 
second LAT catalog (2FGL)
\citep{2012ApJS..199...31N,2012yCat..21990031N} (for Centaurus B see also \cite{2013A&A...550A..66K})
and a Pictor A identification has been discussed in \cite{pictora}.

In the absence of predictions for the $\gamma$-ray luminosity function, 
we follow a phenomenological approach to relate the $\gamma$-ray luminosity to the radio luminosity, 
 as it is commonly done in literature for source
populations  and notably for radio galaxies with the 1FGL data set \cite{inoue}. We explore here for the first time 
the correlation between the {\it core} radio and the $\gamma$-ray luminosity, and adopt  a radio luminosity function from the literature.
The latter is phenomenologically much better established, given the  number of detected MAGN in the radio frequencies should be high. 
A possible correlation between radio and $\gamma$-ray luminosities has been proposed for blazars using the 
Energetic Gamma Ray Experiment Telescope (EGRET) data \citep{padovani1993,stecker1993,salamon1994,dondi1995,narumoto2006}.
Recently, the connection between radio and $\gamma$-ray fluxes has been explored for both the FSRQs and BL Lacs detected by {\it Fermi}-LAT 
during its first year of operation \citep{Ackermann:2011bg,ghirlanda2011}.
On a similar basis, the relation  between radio
emission and  $\gamma$-ray data has been studied for three FRI galaxies observed by
EGRET \citep{ghisellini2005},  as well as for  FRI and FRII galaxies  with 15 months of data taken with {\it Fermi}-LAT  \citep{inoue,misaligned}. 
Variability studies for FRI galaxies support the hypothesis of the compactness of the $\gamma$-ray source \citep{misaligned,2012arXiv1205.1686G}, 
even if a non-negligible $\gamma$-ray counterpart in radio lobes has been observed in Centaurus A \citep{centaurus_a}.
The situation for the FRII population is less definite. A recent {\it Fermi}-LAT analysis of the FRII 3C 111 galaxy \citep{2012ApJ...751L...3G},
together with a multi-frequency campaign conducted in the same period, 
localizes the GeV photons from 3C 111 in a compact, central region associated with the radio core.  

The main radio and $\gamma$ parameters of all the MAGN observed by {\it Fermi}-LAT are reported in Table \ref{tab:galaxies}.
The radio data have been chosen to be the closest in time to {\it Fermi}-LAT data taking. 
Whenever a significant variability has been found, we have selected radio data as
contemporary as possible as the $\gamma$-ray observations.
 Radio data have been taken with the Very Large Array (VLA) for all the objects except NGC 6251, measured with the 
Very Long Baseline Interferometer (VLBI). 
The linear size scales explored by the instruments depend on the redshift of the sources. In our sample, it varies from
about 0.01 kpc to a few kpc \footnote[1]{3C 78: 2 kpc, 3C 274: 40 pc, Cen A: 20 pc, 
NGC 6251: 2 pc, Cen B: 0.5 kpc, For A: 0.1 kpc, 3C 120: 2.3 kpc, PKS 0625-35: 10 pc, Pictor A: 6.9 kpc, 
3C 111: 0.38 kpc, 3C207: 2.8 kpc, 3C 380: 73 kpc.
The radio measurements of some objects (e.g. 3C 207, 3C 280 etc) 
might be contaminated by the extended jet emission. 
However, the uncertainty introduced by this likely contamination is one of the 
uncertainties contributing to the scatter of Fig. 1 and as such is factored in our analysis.},
 except for NGC 6251  and 3C 380. 
Data for \mbox{3C 380} are taken from Effelsberg observations. However, this source 
shows a compact steep spectrum radio morphology and the radio flux from the central region is close to the total
emission and to the flux measured with a few arc sec scale  \footnotemark[1].
 \footnotetext[2]{http://3crr.extragalactic.info/cgi/database}
For 3C 84 the variability is very pronounced and we have therefore excluded it from our correlation analysis.
 IC 310 lacks measurements of the core at 5 GHz, and the total radio flux is very faint.
 For PKS 0943-76 only upper limits for the core are given. For these reasons, these two galaxies 
 are listed but not considered in our analysis
\footnote[3]{Making use of Eq. \ref{eq:core_tot_lara} we can estimate a core radio 
luminosity for IC 310 and PKS 0943-76 which is in agreement with Eq. \ref{eq:correlation}}.
The  photon index $\Gamma$ valid between 0.1 and 10 GeV, and the {\it
  Fermi}-LAT flux integrated for E$_\gamma > $0.1 GeV have been taken from \cite{misaligned} for 3C 78, 3C 111, 3C 120 and from \cite{2012yCat..21990031N} for the remaining objects.
From Table \ref{tab:galaxies}, the mean photon index  $\Gamma$ is 2.37, with spread 0.32.
These numbers are consistent with the values indicated by \cite{inoue}.
We notice that the power-law spectral slope is similar to the one of both blazars, 
$2.40\pm 0.02$ \citep{2010ApJ...720..435A}, and the diffuse  $\gamma$-ray background, $2.41\pm0.05$ \citep{IDGRB}.

\begin{sidewaystable}[p]
\caption{Main radio and $\gamma$ properties of the MAGN observed by {\it Fermi}-LAT. 
Column 1: name of the MAGN (radio classification: FRI or FRII), 2:
redshift, 3: Galactic latitude, 4: spectral index for radio core (total) spectrum 
in a range including 5 GHz, 5: measured radio core (total)  flux at 5 GHz; 
6: photon index for $\gamma$-ray spectrum between 100 MeV - 100 GeV; 7-$\gamma$-ray flux above 100 MeV;
8- Radio core luminosity at 5 GHz ; 9- $\gamma$-ray luminosity
 \\
 References: 1-\cite{morganti}; 2-\cite{pauliny}; 3-\cite{ekers}; 4-\cite{kuehr}; 5-\cite{spinrad}; 
6-\cite{mullin}; 7-\cite{nagar}; 8-\cite{giovannini}; 9-\cite{laing}; 10-\cite{israel}; 11-\cite{burns};
12-\cite{wright}; 13-\cite{evans}; 14-\cite{mantovani}; 15-\cite{jonescenb}; 16-\cite{massardi}; 17-\cite{geldzahler}; 
18-\cite{becker}; 19-\cite{perley}; 20-\cite{linfield}; 21-\cite{kadler}; 22-\cite{burgess}; 23-\cite{kadler}; 24-\cite{gregory91}
}
\vspace{0.5cm}
\label{tab:galaxies}
\centering
\begin{tabular}{|c|c|c|c|c|c|c|c|c|} 
    \hline 
MAGN(FRI,FRII) & $z$ & $b$ [$\degr$] & $\alpha_{\rm core}(\alpha_{\rm tot})$ & $S_{\rm core}^{\rm 5GHz}$ [Jy] ($S_{\rm tot}^{\rm 5GHz})$ [Jy]) & $\Gamma$ & $F_{\gamma}$ [$10^{-9}$ph cm$^{-2}$ s$^{-1}$]  & $L_{r,\rm core}^{\rm 5GHz}$  [erg s$^{-1}$] & $L_{\gamma}$ [erg s$^{-1}$] \\
\hline  
 3C 78/NGC 1218(I)   & 0.0287 & -44.6 &  0 (0.64 $^1$) & $0.964\pm 0.164^1$ ($3.40\pm0.11$$^2$)  & $1.95\pm0.14$ & $4.7\pm1.8$ & (8.8 $\pm$ 1.4 $) \cdot 10^{40}$ & (1.11 $\pm$ 0.54 $) \cdot 10^{43}$
\\
3C 274/M 87(I) & 0.0038 & 74.5 & 0 (0.79 $^8$) & $3.0971 \pm 0.0300$$^7$ ($71.566\pm0.993$ $^{9}$)&$2.17\pm0.07$ & $25.8\pm3.5$ & (4.90 $\pm$ 0.05 $) \cdot 10^{39}$ & ( 6.2 $\pm$ 1.1 $) \cdot 10^{41}$
\\
Cen A(I)  & 0.0009  & 19.4 & 0.30$^{10}$ (0.70$^{10}$) & $6.984\pm0.210$$^{11}$ ($62.837\pm0.099$ $^{12}$) & $2.76\pm0.05$ & $175\pm10$ & (6.19 $\pm$ 0.19 $) \cdot 10^{38}$ & (1.14 $\pm$ 0.09 $) \cdot 10^{41}$
\\
NGC 6251(I) & 0.0247 & 31.2 & 0(0.72$^{9}$) & $0.38\pm0.04$ $^{13}$ ($0.510\pm0.050$ $^{13}$ \footnotemark[1]) & $2.20\pm0.07$ & $18.2\pm2.6$ & (2.57 $\pm$ 0.27 $) \cdot 10^{40}$ & (1.82 $\pm$ 0.41 $) \cdot 10^{43}$
\\
Cen B(I) & 0.0129 & 1.68 & 0 (0.13$^{16}$)   & 2.730$^{15}$ ($6.58\pm 1.04$$^{16}$)  & $2.33\pm0.12$ & $39.3\pm11.4$ & 5.02 $\cdot 10^{40}$ & (8.6 $\pm$ 3.2 $) \cdot 10^{42}$
\\    
For A(I) & 0.00587 & -56.7 & 0.50$^{17}$(0.52 $^1$) & 0.051$^{17}$ (72$^1$)  & $2.16\pm0.15$ & $7.7\pm2.4$ &  1.93 $\cdot 10^{38}$ & (4.6 $\pm$ 2.2$) \cdot 10^{41}$
\\
3C 120(I) & 0.0330 & -27.4  & 0 (0.44$^{18}$) & $3.458\pm0.588$ $^1$ $(8.60\pm1.46$$^1$) & $2.71\pm0.35$ & $29\pm17$ & (4.20 $\pm$ 0.71 $) \cdot 10^{41}$ & (2.9 $\pm$ 1.6 $) \cdot 10^{43}$
\\ 
PKS 0625$-$35(I)$ \footnotemark[2]$ & 0.0546 & -20.0 & 0 (0.65 $^3$) & $0.600\pm0.030$ $^3$ $(2.25\pm0.09$  $^4$)  & $1.93\pm0.09$ & $12.9 \pm 2.6$ & (2.02 $\pm$ 0.10 $) \cdot 10^{41}$ & (1.21 $\pm$ 0.43 $) \cdot 10^{44}$
\\
Pictor A(II) & 0.0351 & -34.6& 0 (1.07 $^1$) & $1.15\pm0.05$$^{19}$ ($15.45\pm0.47$$^4$) & $2.93\pm0.03$ & $21.9\pm3.6$ & (1.58 $\pm$ 0.07 $) \cdot 10^{41}$ & (2.13 $\pm$ 0.46 $) \cdot 10^{43}$
\\ 
3C 111(II) & 0.0485 & -8.61 & -0.20   \footnotemark[2] (0.73$^5$) & 1.14$^{20}$ ($6.637\pm0.996$$^{18}$) & $2.54\pm0.19$ & $40\pm8$ & 2.98 $\cdot 10^{41}$  & (1.01 $\pm$ 0.38 $) \cdot 10^{44}$ 
\\
3C 207(II))$ \footnotemark[3]$ & 0.681 & 30.1 & 0 (0.90 $^5$) & $0.5391\pm0.0030$$^6$ ($1.35\pm0.04$ $^4$)  & $2.36\pm0.11$ & $17.3\pm3.3$ & (3.32 $\pm$ 0.02 $) \cdot 10^{43}$ & (2.41 $\pm$ 0.61 $) \cdot 10^{46}$
\\   
3C 380(II))$ \footnotemark[3]$ & 0.692 & 23.5 &  0 (0.71$^{9}$)  & $5.073\pm0.105$ $^{14}$ ($7.45\pm0.37$ $^4$) & $2.34\pm0.07$ & $30.3\pm3.7$ & (3.12 $\pm$ 0.07 $) \cdot 10^{44}$ & (4.44 $\pm$ 0.73 $) \cdot 10^{46}$\\
\hline
IC 310(I) & 0.0189 & -13.7 & n.a.(0.75$^{23}$) &  n.a. ($0.258\pm0.031$$^{24}$)    & $2.10\pm0.19$ & $11.1\pm6.2$ & - & (7.9 $\pm$ 4.9 $) \cdot 10^{42}$
\\
3C 84/NGC 1275(I)   &  0.0176 & -13.2 & (0.78$^5$) & high variability & $2.00\pm0.02$ & $175\pm8$ & - & (1.22 $\pm$ 0.07 $) \cdot 10^{44}$
\\ 
PKS 0943$-$76(II) & 0.270 & -17.2 & n.a. & upper limits(0.757$^{22}$) & $2.44\pm0.14$ & $19.5\pm5.1$ & - & (2.47 $\pm$ 0.71 $) \cdot 10^{45}$
\\ 
\hline 
\end{tabular}
\footnotetext[1]{VLBI core+jet data, used in our analysis}
\footnotetext[2]{our interpolation}
\footnotetext[3]{non-standard}
\end{sidewaystable}

The  $\gamma$-ray luminosity between energies $\epsilon_1$ and $\epsilon_2$ is given by:
    \begin{equation}
     \label{Lg}
         L_{\gamma}(\epsilon_1, \epsilon_2) =4\pi d^2_L(z) \frac{S_{\gamma}(\epsilon_1, \epsilon_2)}{(1+z)^{2-\Gamma}},
    \end{equation}
where $d_L(z)$ is the luminosity distance at the redshift $z$ and $S(\epsilon_1, \epsilon_2)$ 
is the observed energy flux between $\epsilon_1$  and $\epsilon_2$. 
The factor $(1+z)^{2-\Gamma}$ is the so-called K-correction term that takes into account 
the redshift modification between the emitted and observed energies.
The energy flux $S_{\gamma}(\epsilon_1, \epsilon_2) $
 is linked to the photon flux $F_{\gamma}=\int^{\epsilon_{2}}_{\epsilon_1}d\epsilon \,dN/d\epsilon$ 
(in units of photons cm$^{-2}$ s$^{-1}$) by the  relation:
    \begin{equation} 
    \label{Sg}
        S_{\gamma}(\epsilon_1, \epsilon_2) = \int^{\epsilon_{2}}_{\epsilon_1} \epsilon \frac{dN}{d\epsilon} d\epsilon\, ,
    \end{equation}
where $dN/d\epsilon$ is the $\gamma$-ray spectrum of the source. 

Spectra for the sources in Table \ref{tab:galaxies} have been taken from the 2FGL.
They are simple power-law or log-parabola spectra:
\begin{equation} 
    \label{dNde}
        \frac{dN}{d\epsilon}  = K\left( \frac{\epsilon}{\epsilon_{\rm Pivot}} \right)^{-\Gamma-\beta \log {\left( \epsilon/\epsilon_{\rm Pivot} \right)}}\, , 
\end{equation}
where K is a normalization factor and the parameter $\beta$ being zero for a power-law spectrum\footnotemark[1].
 \footnotetext[1]{(indeed for all the sources considered 
 in our analysis $\beta$=0, as indicated in the 1FGL and 2FGL catalogs. The only source better reproduced by a log-parabola
is 3C84, which is not included in our analysis.)}
 Throughout the paper $\epsilon_1 = 0.1$ GeV, $\epsilon_2 = 100$ GeV, 
   while $\epsilon_{\rm Pivot}$ has been varied for each source except when dealing with average properties ($\epsilon_{\rm Pivot}$=0.1 GeV). 
   \\
Radio luminosity is calculated for a fixed frequency following: 
    \begin{equation} 
    \label{Lr}
         L_{r}(\nu) =\frac{4\pi d^2_L(z)}{(1+z)^{1-\alpha_r}} \; S_{r}(\nu)
    \end{equation}
where $\alpha_r$ is the radio spectral index ($\alpha_{\rm core}$ or $\alpha_{\rm tot}$), $\Gamma=\alpha_r+1$
    and $S_{r}(\nu)$ is the radio energy flux at a given frequency.     

\begin{figure}
\includegraphics[width=\columnwidth]{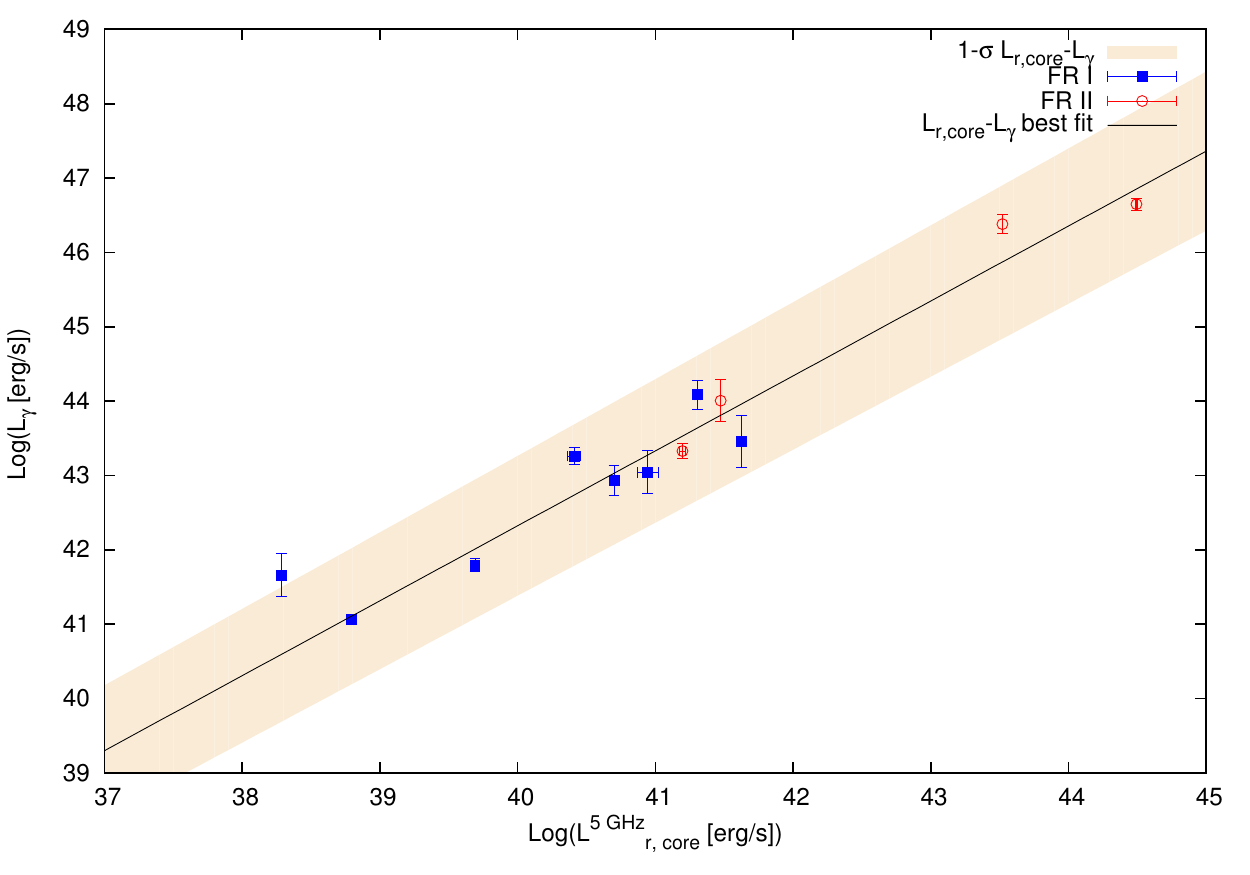}
\caption{ Observed $\gamma$-ray luminosity vs radio core luminosity at 5 GHz for the MAGN of Table \ref{tab:galaxies}. 
Blue squares (red open circles) correspond to possible FRI (FRII) classifications. The solid black line represents the 
calculated correlation as in Eq.\,\ref{eq:correlation}. The light pink shaded area takes into account
the 1$\sigma$ error band in the derived correlation function.}
\label{fig:correlation} 
\end{figure}

\begin{figure}
\includegraphics[width=\columnwidth]{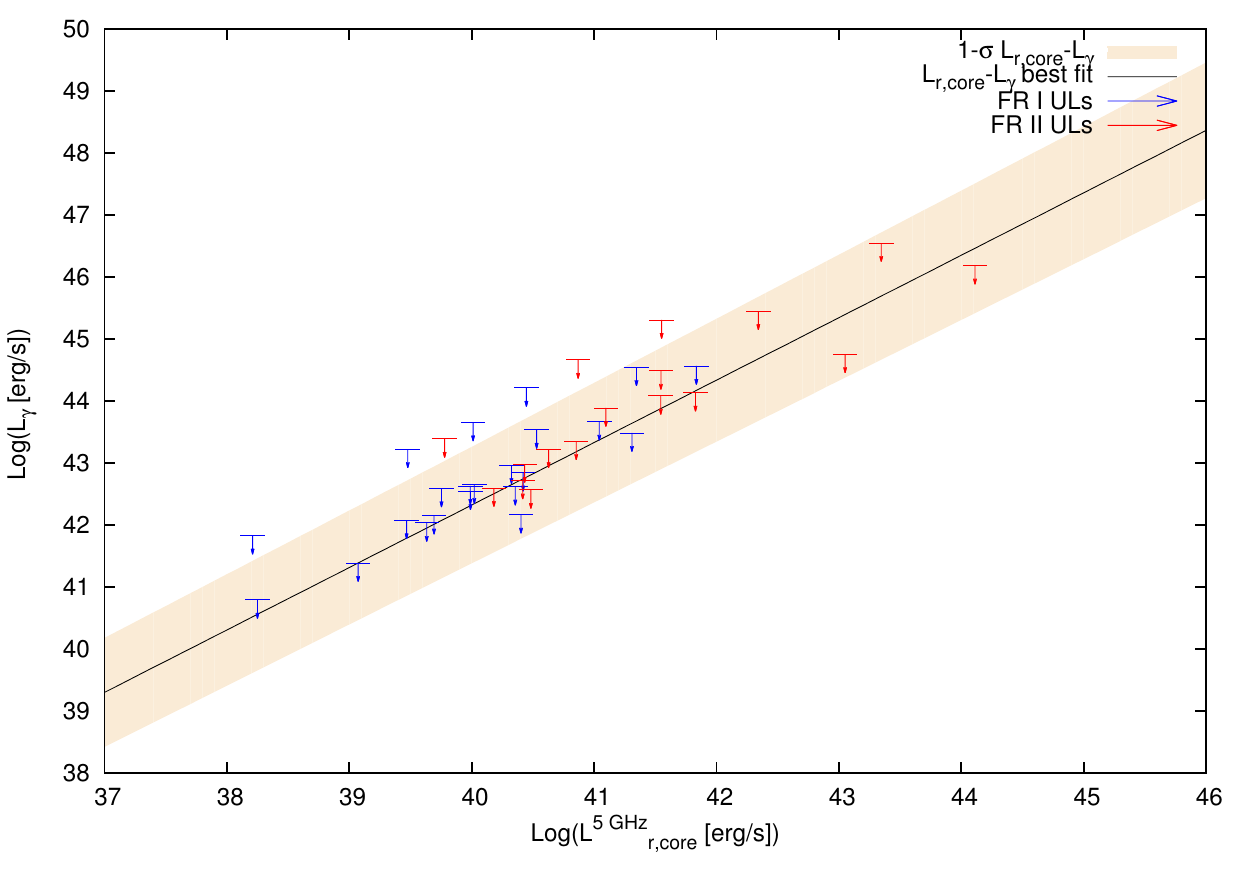}
\caption{Upper limits on several  {\it Fermi}-LAT undetected radio-loud MAGN. Blue (red) symbols refer to possible FRI (FRII) classification. The correlation in Eq. \ref{eq:correlation} (solid black line) is displayed together with the 1$\sigma$ error band (light pink shaded area).}
\label{fig:upper_limits} 
\end{figure}

In Fig. \ref{fig:correlation}  we plot the core radio and $\gamma$ luminosities for the first 12 MAGN listed in Table \ref{tab:galaxies}
(the last three have been excluded from the sample as explained above). 
The possible classification into FRI and FRII RGs is also displayed.
We have calculated luminosities according to Eqs. \ref{Lg}-\ref{Lr},  propagating errors on $\Gamma$ while neglecting errors 
on the redshift, given their negligible effect .
The correlation between $L_{r, {\rm core}}$ and $L_\gamma$  for the 12 objects is described by the function:
 \begin{equation}
     \label{eq:correlation}
         \log{(L_{\gamma})} = 2.00\pm0.98 + (1.008 \pm0.025)\log{ (L^ {5 {\rm GHz}}_{r, {\rm core}} )} \, ,
    \end{equation} 
represented by the solid line in Fig. \ref{fig:correlation}, while the relevant $1\sigma$ error band is shown as a shaded area. 
 It is obtained from the errors on both the $\gamma$-ray and radio luminosities, but the former dominates the uncertainty.  
Eq. \ref{eq:correlation} describes a linear correlation in the log-plane with a coefficient very close to one.
$\gamma$-ray  luminosities are greater than radio luminosities by about two orders of magnitude. The uncertainty
band of the $\gamma$-ray fluxes measured by the {\it Fermi}-LAT spans one order of magnitude
around the best fit. The significance of the correlation is tested in Sec.\ref{sec:testcorr}.

It is useful to compare the $L_{r, {\rm core}}$ - $L_\gamma$ correlation obtained by
 removing three sources with non standard properties from the set of 12 MAGNs in Table\ref{tab:galaxies}. 
The radio morphology of 3C 380 shows a clear core-jet structure when observed by VLBI, 
while at lower resolution it is sorrounded by a steep-spectrum low-surface brightness halo
\citep{misaligned,1991MNRAS.248...86W}. The FRII 3C 207 behaves as a steep spectrum radio quasar in the optical band \citep{misaligned}, 
while PKS 0625$-$35 has no clear association.
  We have therefore also calculated the correlation function excluding the galaxies 3C 380, 3C 207 and PKS 0625$-$35:
 \begin{equation}
     \label{eq:correlation9}
         \log{(L_{\gamma})} = 2.1\pm 2.1 + (1.005 \pm0.055)\log{ (L^ {5 {\rm GHz}}_{r,{\rm core}} )}.
    \end{equation} 
The result is not very different from Eq. \ref{eq:correlation}, if it were not for the increased spread 
in the fitted coefficients. Indeed, 3C 207 and PKS 0625$-$35 have large errors and 3C 380 is quite close to the correlation with the whole sample.  

We report here also the correlation between the total radio luminosity at 5 GHz and the $\gamma$-ray luminosity for the whole sample of 12 sources. The correlation is found to be:
\begin{equation}
     \label{eq:correlationtot}
         \log{(L_{\gamma})} =  -2.5 \pm  1.1+ (1.095 \pm 0.026)\log{ (L^ {5 {\rm GHz}}_{r,{\rm tot}} )} \,. 
 \end{equation} 
The experimental values for the total radio luminosity are quoted in Table \ref{tab:galaxies}.  
The fit for the sample of 9 sources results with:
\begin{equation}
     \label{eq:correlationtot9}
         \log{(L_{\gamma})} =  3.5 \pm  2.3+ (0.948 \pm 0.056)\log{ (L^ {5 {\rm GHz}}_{r,{\rm tot}} )} \,. 
 \end{equation} 
The correlation implied by Eq.\,\ref{eq:correlation} is close to the one 
obtained in \cite{ghisellini2005} for the very small sample of three EGRET $\gamma$-ray loud FRI galaxies
(moreover, one of the three galaxies is Centaurus A,
which might have a non negligible lobe contribution).
In the case of blazars the slope of the correlation between $L_{\gamma}$($>$ 100 MeV) and radio luminosity at different frequencies was found to be: 
1.07$\pm$ 0.05 at 20 GHz \citep{ghirlanda2010}, 1.2 $\pm$ 0.1 at 5 GHz \citep{stecker1993}  and 1.06 $\pm$ 0.02 at 8.4 GHz \citep{yan2012new}. 
The slope coefficient of the correlation for RGs is therefore similar
to the  correlation for blazars. This might indicate that the $\gamma$-ray
emission mechanism is  similar for MAGN and blazars.  We therefore assume that the correlation in Eq. \ref{eq:correlation}
 is a good representation of the luminosity of the cores of MAGN 
and we will employ it in the remainder of this work in order to derive the emission of the MAGN population not detected by the {\it Fermi}-LAT,
but potentially providing a non-negligible diffuse flux.

\section{Upper limits from radio-loud FRI and FRII galaxies not detected by {\it Fermi}-LAT}
\label{sec:UL}
In order to test the  robustness of the core radio-$\gamma$ correlation found in 
Eq. \ref{eq:correlation} we study a sample of radio-loud FRI and FRII 
galaxies that have not been detected by {\it Fermi}-LAT. For these objects 
we derive 95$\%$ C.L. $\gamma$-ray upper limits and verify that they are consistent 
with Eq. \ref{eq:correlation}, given the uncertainty band 
shown in Fig. \ref{fig:correlation}.  The sample has been extracted from RGs in
 \cite{Kataoka2011} and  \cite{ghisellini2005} \
  (first and second block in Table \ref{tab:UL}, respectively), and  represents the sources with the highest radio core fluxes at 5 GHz. 
 Further selection criteria have been applied in defining the sample for our purposes.
From the sample of broad line RGs whose upper limits have been presented in 
 \cite{Kataoka2011}  we have excluded Pictor A \citep{pictora}, detected in the meanwhile, 
 and the sources that do not show a clear FRI or FRII radio morphology classification (RGB J1722+246 and  PKS 2251+11 being 
 Seyfert galaxies, S5 2116+81 being a flat spectrum radio source with a radio jet morphology).
Moreover, sources with latitudes below $10\,^{\circ}$ have been rejected in order to avoid a strong 
contamination from the Galactic plane foreground. This criterion applies to 4C 50.55 ($b = 0.39^{\circ}$).
The same criteria have been applied to sources in 
 \cite{ghisellini2005} leading to the exclusion of 3C 84, 3C 274, 3C 78 already detected 
 in $\gamma$ rays, and 3C 75 that has an atypical RG morphology. 
Finally, 3C 317 has been excluded because of its variability \citep{3C317variability}.
Four FRII RGs from the 3CRR catalogue (3C 245, 3C 109, 3C 212, DA 240) have been added to the sample in order to cover a wider range in radio luminosity  (last block in Table \ref{tab:UL}).  
Our sample is therefore composed of 17 FRII and 22 FRI RGs.
\\
We have computed $\gamma$-ray flux upper limits for the listed galaxies
 by using the {\it Fermi}-LAT Science Tools
  \footnotemark[1].
 \footnotetext[1]{http://fermi.gsfc.nasa.gov/ssc/data/analysis/documentation,
software version v9r27p1, Instrumental Response Functions (IRFs) P7$\_$V6} The
data taking period for the analysis is from the starting time of the mission, 2008
August 4, until 2012 September 9. 
The Mission Elapsed Time (MET) interval runs
from 239557414 to 368928003. Data have been extracted from a region of
interest (ROI) of radius = $8\,^{\circ}$ centered at the position of the
source. This radius represents the best angular region for source analysis as
long as sources are far from the Galactic plane
\citep{2009ApJ...699...31A}, and indeed we neglect in this analysis
sources that lie below $10\,^{\circ}$ in latitude. We selected $\gamma$-rays in the energy range 100 MeV - 100 GeV.
\\
We are using P7SOURCE$\textunderscore$V6 photons. 
Good survey data are selected accordingly to software recommendations, 
with the rocking angle selected to be less than $ 52^\circ$.
Data selection and preparation eliminate photons from the Earth limb by
applying a cut on the zenith angle of $100\,^{\circ}$.
%
An unbinned maximum-likelihood analysis was performed. In the cases where the fit did
not converge we have performed a binned analysis as recommended.
We therefore analyze the source region with both methods and draw the upper limits with the help of the
LATAnalysisScripts \footnote[2]{User contributions
http://fermi.gsfc.nasa.gov/ssc/data/analysis/user/}, which make use of the
UpperLimits.py module.
\\
 Each galaxy in the sample was modeled as a point-like source with a power-law spectrum of index $\Gamma = 2.5$. 
This value has been chosen as nominal spectral index for all MAGN in
analogy with \cite{Kataoka2011}.
We have verified that choosing $\Gamma = 2.3$, closer to the
distribution of the spectral indices from \ref{tab:galaxies} changes the
limits by $\sim 10\%$, while an index of $2.7$ leaves results unchanged.
The number of expected counts in the ROI is derived by considering the emission from all sources in the 2FGL
\footnote[3]{http://fermi.gsfc.nasa.gov/ssc/data/access/lat/2yr$\_$catalog/} 
inside a source region (distance from the target region) of  $13\,^{\circ}$ ($8\,^{\circ}$ + $5\,^{\circ}$). 
The fitting procedure leaves the spectral parameters of all the sources inside the ROI free, 
whereas sources in the region $8^{\circ} < r < 13^{\circ}$ have spectral parameters fixed to the values of the 2FGL.
Additional backgrounds are the Galactic  diffuse emission and the isotropic diffuse model, 
which includes the true IGRB and the residual particle contamination
\footnote[4]{http://fermi.gsfc.nasa.gov/ssc/data/access/lat/BackgroundModels.html}.
The diffuse models used in the analysis are: gal$\textunderscore$2yearp7v6$\textunderscore$v0.fits for the Galactic diffuse model
 and iso$\textunderscore$p7v6source.txt for the isotropic spectral template. 
All the relevant normalizations have been left as free parameters during the fitting procedure. 
The method used to compute the upper limits is a standard profile likelihood. 
A 95\% upper limit (UL) has been computed when the Test Statistic (TS) was less than 25.
In Table\ref{tab:UL} the flux upper limits are quoted together with the TS value for both unbinned and binned analysis. Given the systematic uncertainty 
arising from the different statistical methods, we adopt as \textsl{upper} limit the highest value for the flux bound. These \textsl{conservative} upper limits are 
shown in Fig.\ref{fig:upper_limits} with the luminosity  correlation from Eq. \ref{eq:correlation} overlaid. It is evident that the calculated upper limits
do not fall below the uncertainty band, thus corroborating our core radio and $\gamma$-ray correlation.
\begin{table*}
\caption{Flux upper limits on a sample of MAGN. 
Column 1: name of the MAGN (radio classification: FRI or FRII), 2: redshift, 3: measured radio core flux at 5 GHz [Jy],
4: TS of unbinned analysis, 5: 95$\%$ C.L.\,upper limit from unbinned
analysis on the flux above 100 {\rm MeV} in units of {\rm $10^{-9}$ ph cm$^{-2}$
s$^{-1}$}, 6: TS of binned analysis, 7: 95$\%$ C.L.\,upper limit from
binned analysis on the flux above 100 {\rm MeV} in units of {\rm $10^{-9}$ ph
cm$^{-2}$ s$^{-1}$}; 8: radio core luminosity at 5 GHz in units of erg s$^{-1}$. \\
 References: 1-\cite{morganti}; 2-\cite{1995ApJS..100....1H}; 3-Third Cambridge
Catalogue of Radio Sources 
 ; 4-\cite{dodson2008}; 5-\cite{neff95}; 6-\cite{1992MNRAS.259P..13P};
7-\cite{2002ApJS..141..311T}.
}
\vspace{0.5cm}
\label{tab:UL}
\centering
\begin{tabular}{|c|c|c|c|c|c|c|c|} 
    \hline   
  MAGN(FRI,FRII) & $z$  & $S^{\rm core}_{\rm 5GHz}$ [Jy] &  TS$_{unbinned}$ &
F$_{unbinned}^{UL}$  &  TS$_{binned}$ & F$_{binned}^{UL}$ & $L_{r,\rm core}^{\rm 5GHz}$  \\
    \hline  3C 18 (II)   &0.188 &0.083$^{1}$ & $<$ 1 & 2.7 & 2.6 & 6.0 & 3.51 $\cdot 10^{41}$ \\
    \hline  B3 0309+411B (II)  &0.134 &0.320$^{2}$ & -  & - & $<$ 1 & 5.8 & 6.73 $\cdot 10^{41}$ \\     
    \hline  3C 215 (II)  & 0.412& 0.0164$^{3}$& $<$1 & 3.1 & 4.1 & 6.0 & 3.56 $\cdot 10^{41}$\\
    \hline  3C 227 (II)  & 0.086& 0.032$^{1}$& $<$ 1 & 0.1 & $<$ 1& 1.1 & 2.70 $\cdot 10^{40}$ \\
    \hline  3C 303 (II)   & 0.141& 0.150$^{3}$& $<$ 1& 2.8 &3.3 &4.6 & 3.50 $\cdot 10^{41}$\\
    \hline  3C 382 (II)  & 0.058& 0.188$^{3}$&$<$ 1& 4.1 &1.2& 5.9 & 7.12 $\cdot 10^{40}$\\
    \hline  3C 390.3 (II)  & 0.056& 0.120$^{4}$&$<$ 1& 1.7  &3.0 &4.7 & 4.26 $\cdot 10^{40}$\\
    \hline  3C 411(II)  & 0.467& 0.078$^{5}$& - & - & $<$ 1& 6.1 & 2.2 $\cdot 10^{42}$\\
    \hline  4C 74.26 (II)  & 0.104& 0.100$^{6}$& 1.1 & 5.4 & $<$1&5.7 & 1.25 $\cdot 10^{41}$\\
    \hline  PKS 2153-69 (II)  &0.028 &0.300 $^{7}$& 4.2 & 6.6 &$<$1 &6.2 & 2.67 $\cdot 10^{40}$\\
    \hline  3C 445 (II)  & 0.056& 0.086$^{1}$&$<$1 &0.8  &$<$1 &1.0 & 3.06 $\cdot 10^{40}$\\
    \hline  3C 465 (I)   &0.029 & 0.270$^{3}$& - & - & $<$ 1 & 0.5 & 2.5 $\cdot 10^{40}$ \\
    \hline 
    \hline  3C 346 (I) & 0.162& 0.220$^{3}$& 4.5 & 6.4  & 10.8 & 10.2 & 1.39 $\cdot 10^{39}$\\ 
    \hline  3C 264 (I) & 0.021& 0.200$^{3}$& 9.0& 5.7  &14.0&  7.5 & 9.58 $\cdot 10^{39}$\\
    \hline  3C 66B (I)&0.022& 0.182$^{3}$& - & - & $<$ 1 & 8.3 & 9.31 $\cdot 10^{39}$\\
    \hline  3C 272.1(I)& 0.003& 0.180$^{3}$& 5.2& 5.6 &5.3 &6.8 & 1.66 $\cdot 10^{38}$ \\
    \hline  3C 315 (I)& 0.1083& 0.150$^{3}$& - & - & $<$ 1 & 2.1 & 2.04 $\cdot 10^{41}$ \\
    \hline  3C 338 (I)& 0.030& 0.105$^{3}$& - & - & $<$ 1 & 4.6 & 1.07 $\cdot 10^{40}$\\
    \hline  3C 293 (I)& 0.045& 0.100$^{1}$& $<$ 1& 1.5 &$<$1 &1.8 & 2.29 $\cdot 10^{40}$\\
    \hline  3C 29 (I)& 0.045& 0.093$^{3}$& $<$ 1& 1.5 & $<$1&4.1 & 2.11 $\cdot 10^{40}$\\
    \hline  3C 31(I)& 0.017& 0.092$^{3}$& - & -& $<$ 1& 4.0 & 2.83 $\cdot 10^{39}$ \\
    \hline  3C 310 (I)& 0.054&0.080$^{3}$ & $<$ 1 & 1.2 & $<$1&2.1 & 2.63 $\cdot 10^{40}$ \\
    \hline  3C 296 (I)& 0.024& 0.077$^{3}$& $<$ 1& 1.5 &$<$1 &2.3 & 4.79 $\cdot 10^{39}$\\
    \hline  3C 89 (I)& 0.1386 & 0.049$^{3}$& - & - & $<$1 & 1.8 &  1.10 $\cdot 10^{41}$\\    
    \hline  3C 449 (I)& 0.017& 0.037$^{3}$& $<$ 1& 0.5 &$<$1&0.8 & 1.19 $\cdot 10^{39}$\\
    \hline  3C 288 (I)& 0.246& 0.030$^{3}$&$<$ 1& 1.5 & 1.6&3.7 & 2.22 $\cdot 10^{41}$\\
    \hline  3C 305 (I)& 0.0414 & 0.0295$^{3}$& - & - & $<$1 &2.1 & 5.66 $\cdot 10^{39}$\\
    \hline  3C 83.1B (I)& 0.026& 0.040$^{3}$& 10.0 & 19.7 &16.5 & 23.2 & 2.89 $\cdot 10^{39}$\\
    \hline  3C 424 (I)& 0.1270 &  0.0180 $^{3}$& - & - & $<$1 & 1.6 & 3.39 $\cdot 10^{40}$\\
    \hline  3C 438 (II)& 0.290& 0.0071$^{3}$& $<$ 1& 0.9& $<$1&3.2 & 7.40 $\cdot 10^{40}$\\
    \hline  3C 386 (I)& 0.018& 0.120$^{3}$& -& - &  $<$1 &3.2 & 4.15 $\cdot 10^{39}$ \\
    \hline  3C 277.3 (I)& 0.0857& 0.0122$^{3}$& -& - &  4.2 & 5.1 & 1.03 $\cdot 10^{40}$\\
    \hline  3C 348 (I)& 0.1540& 0.010$^{3}$&- & - & $<$1&5.1 &  2.80 $\cdot 10^{40}$\\
    \hline  3C 433 (II)& 0.102& 0.005$^{3}$&- & - & $<$1&1.9 & 5.96 $\cdot 10^{39}$\\
    \hline  3C 442A (I)& 0.027& 0.002$^{3}$&$<$1 &0.7 & $<$1  & 0.9 & 1.62 $\cdot 10^{38}$\\
    \hline  
    \hline  3C 245 (II)& 1.029& 0.910$^{3}$& $<$ 1& 2.0 & $<$1&4.0 & 1.30 $\cdot 10^{44}$\\
    \hline  3C 109 (II)& 0.306& 0.263$^{3}$&$<$ 1& 1.4& $<$1 &3.5 & 3.06 $\cdot 10^{42}$\\
    \hline  3C 212 (II)& 1.049& 0.150$^{3}$& 6.4& 7.1 & 10.11&8.8 & 2.22 $\cdot 10^{43}$\\
    \hline  da 240 (II)& 0.036& 0.105$^{3}$& $<$ 1& 1.5 & $<$1&2.8 & 1.48 $\cdot 10^{40}$\\ 
        \hline
\end{tabular}
\end{table*}

\section{Test of  the radio-$\gamma$ correlation}
\label{sec:testcorr}

The correlation established in Eq.\,\ref{eq:correlation} could be biased by distance dependence of the luminosity and flux-limited samples
\citep{1992A&A...256..399P,2011MNRAS.413..852G,inoue}.
We have tested the strength of the correlation via a partial correlation analysis, in order to verify that the radio core - $\gamma$-ray luminosity 
correlation for MAGN is not spurious. 

Firstly, we calculate the Spearman rank-order correlation coefficient. The Spearman correlation coefficients are 0.94, 0.92, 0.98 between  
$\log{(L_{ r,{\rm core}}^ {5 {\rm GHz}})}$ and $\log{(L_{\gamma})}$, between $\log{(L_{ r,{\rm core}}^ {5 {\rm GHz}})}$ and 
redshift, and between $\log{(L_{\gamma})}$ and redshift, respectively. The partial correlation coefficient turns out to be 0.51 and the null hypothesis that the two 
luminosities are uncorrelated is rejected at the 95$\%$ C.L. 

\begin{figure}
 \includegraphics[scale=.4]{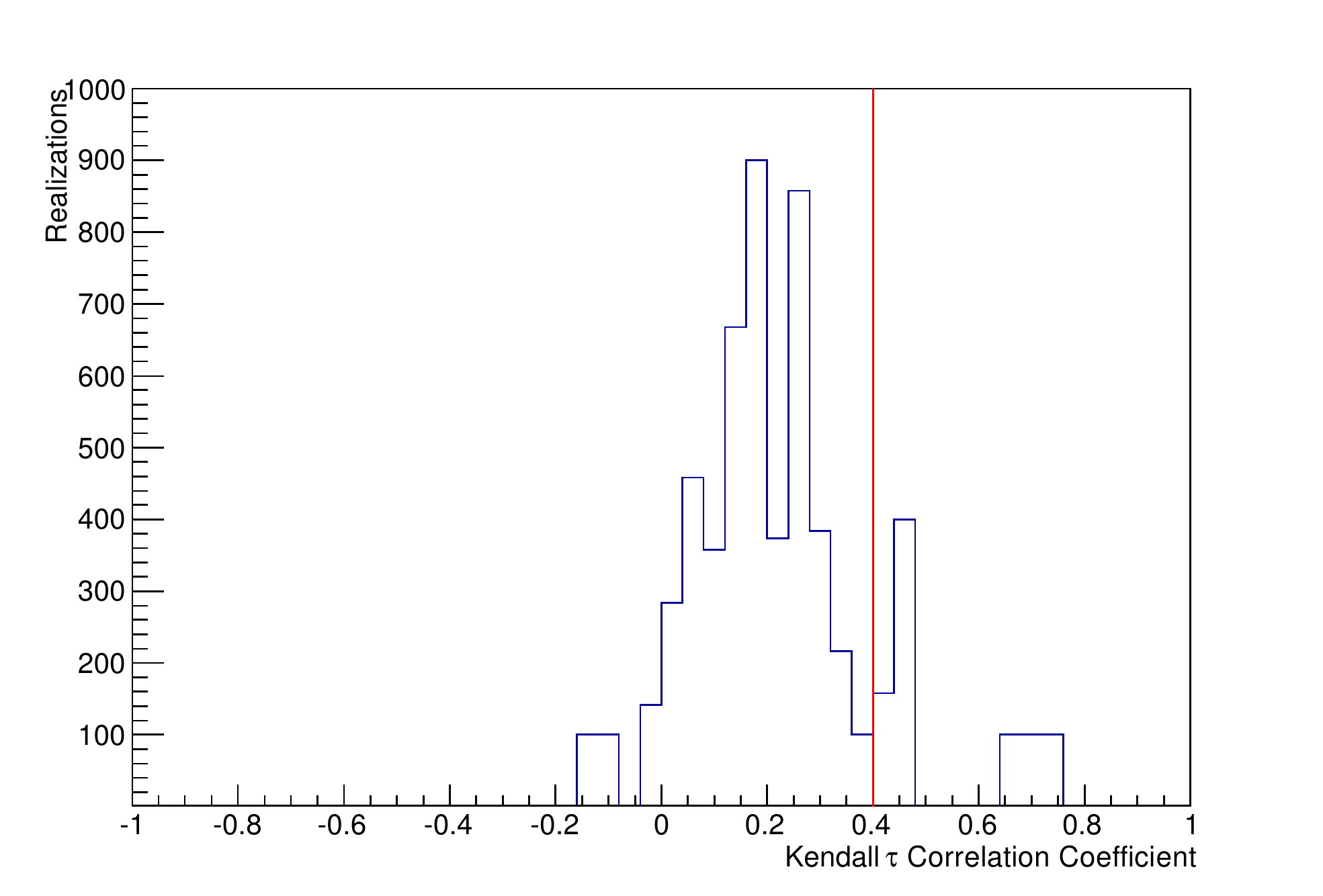}
\caption{ Null hypothesis distribution of $\tau$ correlation coefficients assuming independence between the $\gamma$ and radio wavebands. 
The null hypothesis distributions are 
generated from 5800 permutations of $\gamma$-ray luminosities among the galaxies by requiring that the resultant $\gamma$-ray fluxes 
exceed the flux threshold of 7.7 $\cdot 10^{-9}$ photons cm$^{-2}$ s$^{-1}$. 
The mean value is 0.223 with the standard deviation of the distribution RMS=0.173. 
The correlation coefficient of the actual data is represented by the red solid line, $\tau$ = 0.397.}
\label{fig:kendall} 
\end{figure}

Moreover, we test the significance of the radio-$\gamma$ correlation by using the modified Kendall $\tau$ rank correlation test proposed by \cite{1996MNRAS.278..919A}, 
which is suitable for partially-censored datasets. By following the procedure highlighted in \cite{2012ApJ...755..164A}, we perform a Monte Carlo simulation in order to 
compute the distribution of the $\tau$ correlation coefficients obtained under the null hypothesis of independence between the two wavebands.
Starting from the detected sample of 12 sources we build several dataset realizations by scrambling the derived $\gamma$-ray luminosities among galaxies. 
For each galaxy we then compute the corresponding flux and we retain only galaxies with a flux above the minimal $\gamma$-ray flux of the detected sample
 (7.7 $\cdot 10^{-9}$ photons cm$^{-2}$ s$^{-1}$). 
If the scrambled sample has fewer than 12 sources above the flux threshold, we randomly extract an additional source from the upper limit dataset
(from Table \ref{tab:UL}) until the flux threshold is reached.
For each scrambled dataset we then compute the Kendall coefficient and we build its distribution as shown in Fig. \ref{fig:kendall}. The displayed distribution refers to 5800 
realizations of scrambled samples and the red line represents the value of the $\tau$ correlation coefficient of the actual data, $\tau$ = 0.397.

Finally, we compare the $\tau$ correlation coefficient of the actual data to the distribution of $\tau$ and we find that the integral of the distribution above $\tau$ = 0.397 is 0.05. 
This is the probability to obtain the actual correlation by chance,
\textsl{i.\,e.\,} the p-value of the correlation (the smaller the p-value, the greater the probability for the observed correlation of being true).
As in the case of the Spearman test, we can exclude the correlation happening by chance at the 95$\%$ C.\,L\,.
The result indicates a physical correlation between the core radio emission and the $\gamma$-ray flux of the {\it Fermi}-LAT detected MAGN.

\section{The  $\gamma$-ray luminosity function}
\label{sec:GLF}

The luminosity function for a given energy defines 
the number of sources emitting at that energy per unit comoving volume, per unit (base 10) logarithm of luminosity:
  \begin{equation} 
  \label{rhodef}
        \rho(L,z) = \frac{d^2N}{d\log(L)\;dV}.
    \end{equation} 
In the radio band, data are available for hundreds of radio-loud MAGN,
depending on the frequency of the survey. Usually radio observations refer to the total emission 
of the AGN, including the central region, jets and radio lobes. 
Only for a limited number of objects detected at low radio frequencies  (around 0.1-few GHz), 
the flux from the central core alone has been measured. 
The RLF is derived phenomenologically  by fitting data on the emission of the radio sample.
Results on the total RLF are quite well established  \citep{willott,dunlop,yuan}, 
while the literature about the core radio luminosity function is still limited  \citep{yuan}, 
given the scarcity of experimental data.

\begin{figure}
 \includegraphics[width=\columnwidth]{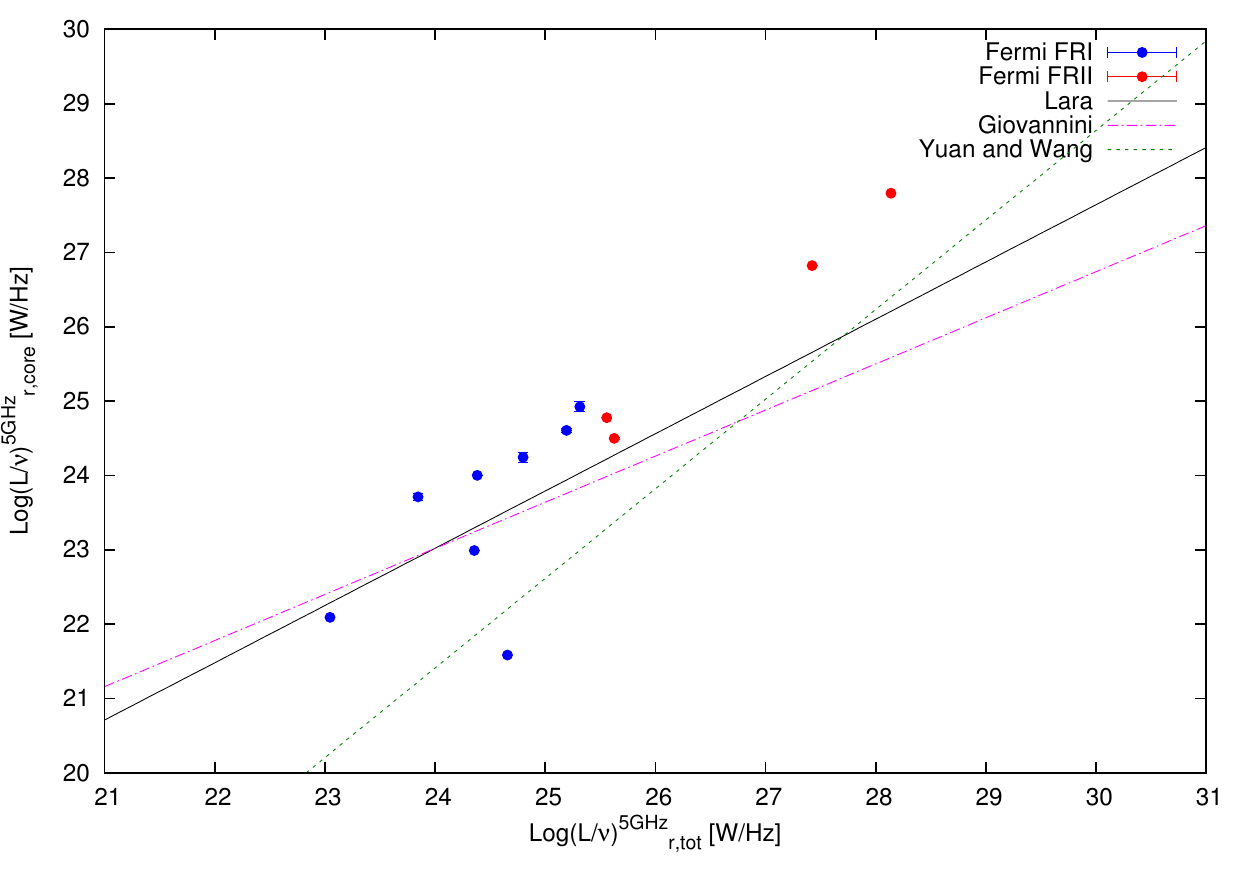}
\caption{Radio core luminosity versus total radio luminosity at 5 GHz. 
Solid black line corresponds to Eq. \ref{eq:core_tot_lara}, 
while the pink dot-dashed and the green dotted correspond to Eq.\ref{eq:core_tot_giovannini} and Eq.\ref{eq:core_tot_yuan}, respectively. 
 Blue squares (red \sout{open} circles) are the experimental data for our sample of FRI (FRII) taken from Table \ref{tab:galaxies}.
 All the points contain error bars, which are indeed very small.}
\label{fig:correlationtotalcore}
\end{figure}

Unfortunately, deriving the GLF from fitting the gamma-ray measurements is not
feasible, due to the small size of the $\gamma$-ray loud MAGN sample. 
Following previous attempts applied to blazars 
\citep{1996ApJ...464..600S,1997ApJ...476....7K,narumoto2006,2011ApJ...736...40S} 
and, to a lesser extent, to 
RGs \citep{ghisellini2005,inoue}, we derive the GLF from the RLF by exploiting the correlation 
between radio and $\gamma$-ray luminosities found in Sect.  \ref{sec:correlation}.
We assume that:
\begin{equation}
\label{eq:k}
 N_{\gamma}=k\;N_r,
 \end{equation}
 where the normalization $k$ takes into account our ignorance  of the number of radio-loud MAGN 
emitting in $\gamma$ rays as well ($N_r$ and $N_{\gamma}$, respectively).
 From Eq.\,\ref{rhodef}, it follows 
    that $N=\int dV \int  \rho(L,z) d\log L$ and therefore the GLF is defined through a RLF by:
    \begin{equation}
     \label{rhogamma}
        \rho_{\gamma}(L_{\gamma},z) = k \; \rho_r(L_r,z) \frac{d\log L_r}{d\log L_{\gamma}}\,.
    \end{equation} 
Given the results of the previous sections, the above equation takes the specific form:
 \begin{equation}
     \label{rhogammanew}
        \rho_{\gamma}(L_{\gamma},z) = k \; \rho_{r,{\rm core}}(L^{5 {\rm GHz}}_{r,{\rm core}}(L_{\gamma}),z) 
        \frac{d\log L^{5 {\rm GHz}}_{r,{\rm core}}(L_{\gamma})}{d\log L_{\gamma}}\,,
    \end{equation}
where $\rho_{r,{\rm core}}$ refers to the radio luminosity function of the cores of the MAGN. 
If our hypothesis of a correlation between the core radio and $\gamma$ emission is  physical, 
as supported by the results on the ULs (see previous section),
we might expect $k$ values not too far from 1. In other words, 
each RG with a bright radio core is expected to emit in the $\gamma$-ray band as well. 
 The correlation between radio and  $\gamma$-ray luminosities
is assumed to be a specific analytical expression, Eq. \ref{eq:correlation}, shown to be 
in very good agreement with the data. In this sense, the scatter in the correlation derives only from 
errors in the experimental data and not in a potential scatter on the luminosity  form.  
The radio luminosity is energetically weaker, according to Eq. \ref{eq:correlation}.
As already noted, the lack of a reliable core RLF from data prevents us from using Eq.\ref{rhogammanew} directly.
The only core RLF in \citep{yuan} finds a strong negative evolution, while it is 
expected that core and lobes should co-evolve with redshift. 
Radio galaxies, as well as lobes that are detected at low frequency, evolve positively and 
there is very little evidence for the presence of radio galaxies with a `switched off' core, 
as it should be if lobes and cores had a different evolution. 
Moreover, the same authors report the positive evolution of radio galaxies and derive a 
correlation between total radio flux and core flux that would yield a positively-evolving core RLF 
with a simple transformation of their total RLF (using their correlation).
We will  therefore  make use of the total RLF and obtain the core RLF through the link between 
total and core radio luminosities. 
\\
As a first ingredient, we need a correlation between radio core and total luminosities. 
In Fig.\ref{fig:correlationtotalcore} we display the correlation between
 ${L^{5 {\rm GHz}}_{\nu,{\rm tot}}}$ and ${L^{5 {\rm GHz}}_{\nu,{\rm core}}}$.
The three curves correspond to: 
\begin{equation}
     \label{eq:core_tot_lara}
         \log{L^{5 {\rm GHz}}_{\nu,{\rm core}} } = 4.2\pm 2.1 + (0.77 \pm0.08)\log{L^{1.4 {\rm GHz}}_{\nu,\rm tot}}\, 
    \end{equation} 
from \cite{lara2004} (black solid curve),     
\begin{equation}
     \label{eq:core_tot_giovannini}
         \log{L^{5 {\rm GHz}}_{\nu,{\rm core}}} = 7.6\pm 1.1 + (0.62 \pm0.04)\log{L^{408 {\rm MHz}}_{\nu,\rm tot}}\,
    \end{equation} 
from \cite{giovannini2001} (pink dot-dashed curve),
\begin{equation}
     \label{eq:core_tot_yuan}
         \log{L^{408 {\rm MHz}}_{\nu,{\rm tot}}} = 7.10\pm 0.90 +  (0.83 \pm0.04)\log{L^{5 {\rm GHz}}_{\nu,{\rm core}}}\,
    \end{equation} 
from \cite{yuan} (green dotted curve). 
We report all the luminosities at 5 GHz, assuming a power-law dependence  $L/ \nu \propto \nu^{-\alpha}$, with $\alpha_{\rm tot}$ = 0.80
for the total radio emission (as assumed, e.g., in \citep{inoue}).
It is clear from Fig. \ref{fig:correlationtotalcore}
 that the experimental data for our MAGN sample are best represented by the 
correlation proposed by \cite{lara2004}. We will therefore adopt Eq.\ref{eq:core_tot_lara} throughout the paper. 
 The possible systematics introduced by this correlation are likely compensated, at least to a good extent, 
by the fit to the cumulative number counts (see the following section). 

The core RLF may be obtained from the total one following the same reasoning as for Eqs. \ref{eq:k}-\ref{rhogamma}:
\begin{equation}
     \label{rhorcore}
        \rho_{ r,{\rm core}}(L_{ r,{\rm core}},z) = \rho_{ r,{\rm tot}}(L_{ r,{\rm tot}},z) \frac{d\log L_{ r,{\rm tot}}}{d\log L_{ r,{\rm core}}},
    \end{equation}
where we expect that the number of MAGN showing core and total emission is almost the same. 
We adopt as the total RLF the one derived in  \cite{willott} (Model C with $\Omega_M$=0) and shift luminosities from 151 MHz to 5 GHz 
according to the power law  explained above. 
We convert the comoving volume to the standard 
$\Lambda$CDM cosmology by using the conversion factor $\eta(z)$:
\begin{equation} \label{eta}
\eta(z) = \frac{d^2V_W/dz d\Omega}{d^2V/dz d\Omega},
\end{equation} 
where $d^2V_W/dz d\Omega$ is the comoving volume element used by \cite{willott}:
\begin{equation} 
\label{Vwillott}
\frac{d^2V_W}{dz\;d\Omega}= \frac{c^3z^2(2+z)^2}{4H_{0,W}^3(1+z)^3},
\end{equation} 
$c$ is the speed of light and $H_{0,W} = 50$ km s$^{-1}$ Mpc$^{-1}$. 
In the cosmological model $\Lambda$CDM the comoving volume element is defined as:
\begin{eqnarray}
 \label{Vlambdacdm}
&&\frac{d^2V}{dzd\Omega}=\\
 &&\frac{c\;{d_L(z)}^2}{H_{0}(1+z)^2\sqrt{(1-\Omega_{\Lambda}-\Omega_{M})(1+z)^2+(1+z)^3\Omega_M+\Omega_{\Lambda}}}. \nonumber
\end{eqnarray} 
\\
We finally obtain the GLF inserting  Eq.\ref{rhorcore} in Eq.\ref{rhogammanew}:
\begin{eqnarray}
     \label{rho_gamma}    
        \rho_{\gamma}(L_{\gamma},z) &= &k \; 
     \rho_{ r,{\rm tot}}\left(L^{5 {\rm GHz}}_{ r,{\rm tot}}(L^ {5 {\rm GHz}}_{ r,{\rm core}}(L_{\gamma})),z\right)  \nonumber \\
       & \cdot & \frac{d\log L^{5 {\rm GHz}}_{ r,{\rm core}}}{d\log L_{\gamma}} \, \frac{d\log L^ {5 {\rm GHz}}_{ r,{\rm tot}}}{d\log L^ {5 {\rm GHz}}_{ r,{\rm core}}}\,.
\end{eqnarray}
The ${d\log L^{5 {\rm GHz}}_{ r,{\rm core}}}/{d\log L_{\gamma}} $ will be computed from Eq.\ref{eq:correlation}, while the 
${d\log L^ {5 {\rm GHz}}_{ r,{\rm tot}}}/{d\log L^ {5 {\rm GHz}}_{ r,{\rm core}}}$ derives from the total-core correlation, Eq.\ref{eq:core_tot_lara}. 


\section{Predictions for the source count distribution}
\label{sec:Ncount}
An important observable for the correctness of our method is provided by the source count distribution of MAGN 
measured by {\it Fermi}-LAT. The source count distribution, known also as $\log N-\log S$, is the cumulative number of sources 
$N(>F_{\gamma})$ detected above a threshold flux $F_\gamma$. 
We have derived the experimental source count distribution of the 12 MAGN of our sample following \citep{2010first}:
     \begin{equation} \label{ncountexp}
         N(>F_{\gamma}) = \sum^{N(>F_{\gamma,i})}_{i=1}\frac{1}{\omega(F_{\gamma,i})},
    \end{equation}
where the sum runs on all the $i$-sources with a $\gamma$-ray flux $F_{\gamma,i} > F_{\gamma}$,
and $\omega(F_{\gamma,i})$ is the flux dependent detection efficiency compatible with our sample.
As shown in  \cite{abdo2009,2010first}, at faint fluxes the {\it Fermi}-LAT more easily detects hard-spectrum sources rather than
sources with a soft spectrum.  Sources with a photon index of $\Gamma$=1.5 can be detected down to fluxes that are a factor
$>20$ fainter than those of a source with a photon index of 3.0. Given this strong selection
effect, the intrinsic photon-index distribution is necessarily different from the observed one. This effect is taken into
account by the detection efficiency.  Since the latter is not available for the MAGN sample, 
we reasonably assume it is the same as for blazars of the 1FGL and  take it from  \cite{2010ApJ...720..435A}.
We demonstrate in the Appendix  that an empirical estimation of the efficiency for the 2FGL blazars 
does not change the results of our analysis.
\\
 The theoretical source count distribution $N_{\rm th}(>F_{\gamma})$ for a $\gamma$-ray flux $F_{\gamma}$ 
    is calculated  following the definition of GLF in Eq.\,\ref{rhodef}:
    \begin{eqnarray}
    \label{Ncount}
    N_{\rm th}(>F_{\gamma}) &=&  4\pi \; \int^{\Gamma_{min}}_{\Gamma_{max}} \frac{dN}{d\Gamma}  d\Gamma \int^{z_{max}}_0 \frac{d^2V}{dz d\Omega} dz\nonumber\\
   && \int^{L_\gamma^{max}}_{L_\gamma({F_\gamma},z,\Gamma)}  \; \frac{dL_{\gamma}}{L_{\gamma} \ln(10)} \; \rho_{\gamma}(L_{\gamma},z, \Gamma),
    \end{eqnarray} 
where $L_{\gamma} (F_{\gamma},z, \Gamma)$ is the $\gamma$-ray luminosity of a RG at 
redshift $z$, whose photon spectral index is $\Gamma$ and photon flux is  $F_{\gamma}$  (integrated above 100 MeV). 
The spectral index distribution, $dN/d\Gamma$, is assumed to be Gaussian in analogy with blazars  \citep{2010ApJ...720..435A}. 
Indeed, any observing instrument with finite sensitivity, and {\it Fermi} is no exception to this, inevitably selects sources near its 
detection threshold, resulting in an asymmetric distribution of observed spectral indices. 
The detected MAGN spectral index distribution, similarly to the one from blazars, appears as non-Gaussian and asymmetric (more hard sources than soft sources). 
A proper test, including selection effects, requires a larger sample and is beyond the scope of this paper. On the other hand, 
there are no indications that support deviations from a standard symmetric Gaussian distribution.
Our treatment of the distribution in photon indices does not  explicitly correct for errors in individual measurements, which can  
artificially increase the distribution spread \citep{2007ApJ...666..128V}. 
However, since the errors in individual photon index  measurements are quite small, we expect this effect not to be very  
important. Its effect might slightly decrease the expected emission at  
high energies, which would further reduce the importance of any  
cascade emission for this component of the IGRB, which we do not  
calculate here.
The comoving volume, $ d^2V/(dz d\Omega)$, is computed according to Eq.\ref{Vlambdacdm}.
We fix  $\Gamma_{min}$=1.0, $\Gamma_{max}$=3.5,  $z_{max} = 6$ and $L_{\gamma, max} = 10^{50}$ erg s$^{-1}$.  
 \begin{figure*}
 \begin{centering}
 \includegraphics[]{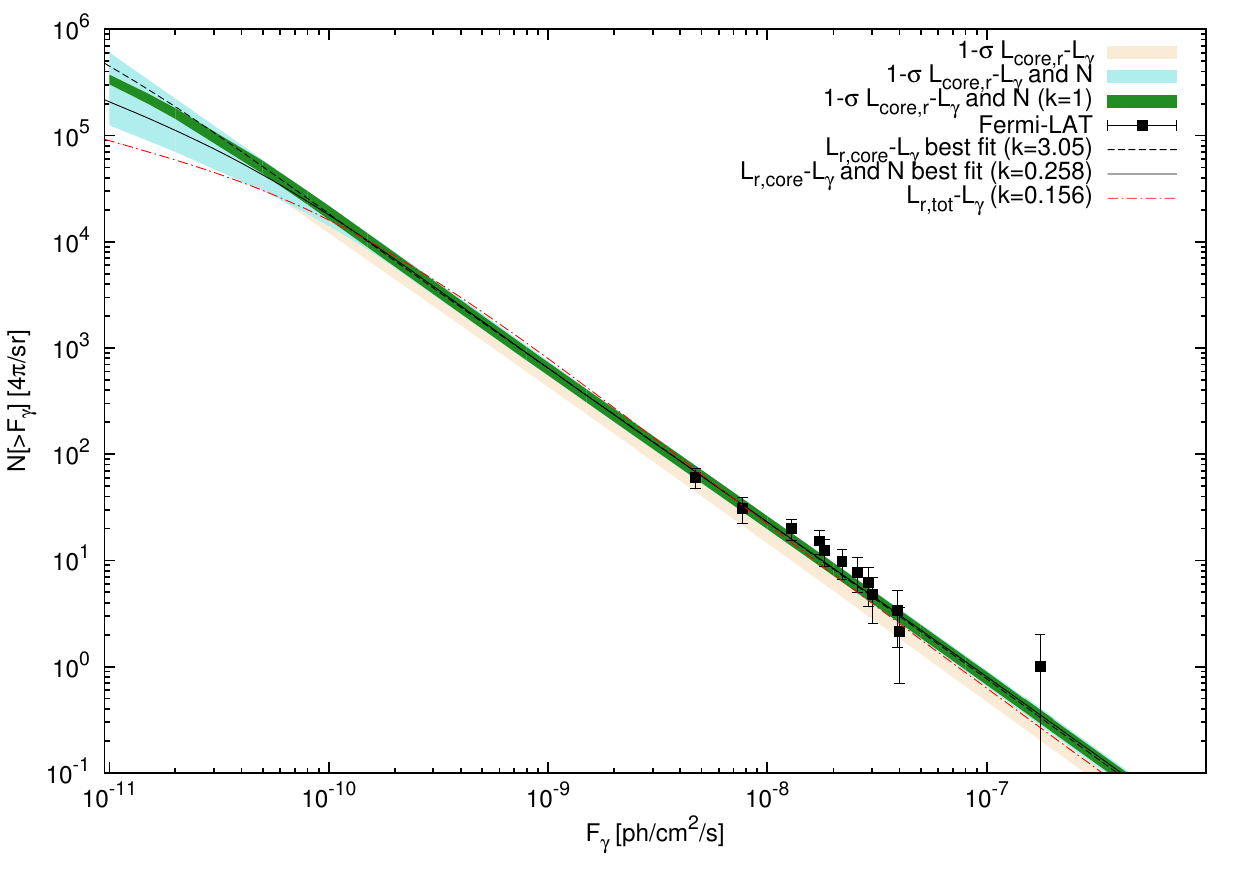}
\caption{ Source count distribution as a function of the integrated $\gamma$-ray flux. 
{\it Fermi}-LAT data are represented by black squares with 1$\sigma$ error bars. 
The black dashed line (pink shaded area) shows the source count distribution predicted with 
the best fit configuration (1$\sigma$ uncertainty band) for the $L_{r, {\rm core}}-L_\gamma$ correlation 
$  \log{(L_{\gamma})} = 2.00\pm0.98 + (1.008 \pm0.025)\log{ (L^ {5 {\rm GHz}}_{r, {\rm core}} )} $. 
The black solid line (cyan shaded area) corresponds to the source count distribution predicted
after the minimization  on the  $L_{r, {\rm core}}-L_\gamma$ best fit (1$\sigma$ uncertainty band) correlation and $k$ 
(see text for details). 
The green shaded area includes all the configurations with $k$=1. 
The red dot-dashed curve has been obtained with the  $L_{r, {\rm tot}}-L_\gamma$ correlation according to Eq.\ref{eq:correlationtot}. 
\label{fig:ncount} }
\end{centering}
\end{figure*}

Fig. \ref{fig:ncount} shows the theoretical $N_{\rm th}(>F_{\gamma})$,
calculated from Eq.\ref{Ncount}, with several bands of uncertainty, overlaid with the experimental source count
distribution from Eq.\ref{ncountexp}.
Their comparison is discussed here only as a consistency check of the
validity of the assumptions involved in Eq.\ref{Ncount} and in
particular of the ratio of MAGN emitting in $\gamma$ rays relative to those emitting
in radio-core, i.e. the $k$ parameter in Eq. \ref{Ncount}.
The data points for the experimental $N(>F_{\gamma})$ are in fact highly
correlated, and a fit to those points is not statistically
meaningful.
Nonetheless, it is useful to fit the theoretical $N_{\rm
  th}(>F_{\gamma})$ to the experimental source count
distribution to constrain the only free parameter $k$ in this calculation.
Additionally, the shape of the function predicting $N(>F_{\gamma})$ is essentially driven 
by the radio luminosity density function, and not by
the fit to the experimental source count distribution.
\\
The  black dashed line in Fig. \ref{fig:ncount} has been derived 
from  the best fit parameters of  Eq.\ref{eq:correlation}
($ \log{(L_{\gamma})} = 2.00\pm0.98 + (1.008 \pm0.025)\log{ (L^ {5 {\rm GHz}}_{r, {\rm core}} )} $),
whose fit to the experimental source count distribution gives $k=3.05\pm 0.20$
with a $\chi^2$=6.98 (for 11 degrees of freedom).
This indicates that the best fit radio core-$\gamma$ correlation function slightly
under-predicts the distribution of MAGN observed by {\it Fermi}-LAT.
\\
For obtaining the bands depicted in Fig. \ref{fig:ncount}, we have proceeded as follows: 
\\
$i$) we have calculated the $N(>F_{\gamma})$ for all the  correlation coefficients
 falling in the 1$\sigma$ uncertainty band for the $L_{r, {\rm core}}-L_\gamma$ relationship (Fig.  \ref{fig:correlation});\\
$ii$) for each combination of these coefficients  
we have determined $k$ from the comparison with  the $\log N-\log S$ (pink shaded area); \\
$iii$) the configuration with the lowest $\chi^2$ among all the configurations explored at point $ii$ 
predicts the best $N(>F_{\gamma})$ (black solid line, $k$=0.258); \\
$iv$) all the configurations giving a 1$\sigma$ variation from the
lowest $\chi^{2}$ (minimal $\chi^2$ +3.53) span the cyan shaded area. 
\\
The red dot-dashed curve was obtained for the radio total - $\gamma$-ray luminosity correlation in 
Eq.\ref{eq:correlationtot} and the total RLF in \cite{willott}. This hypothesis leads to a lower number of sources at the lowest 
fluxes. 
The pink shaded area (and similarly the cyan band) is quite narrow because of the 
degree of freedom implied by $k$, which is fitted on the experimental logN-logS for all 
the $\gamma$-ray luminosities falling in the 1$\sigma$ band of Eq. \ref{eq:correlation}.
\\
Finally, the green shaded band was obtained by fixing 
the normalization factor $k$ in Eqs. \ref{eq:k},\ref{rhogamma} equal
to 1, which represents  the ideal situation in which  we predict
that each MAGN has a radio-loud central region emitting in
$\gamma$ rays as well.
We have varied the luminosity correlation in Eq.\ref{eq:correlation}
within its 1$\sigma$ band. The lowest $\chi^2$ is 6.80 (for 10 degrees of freedom), 
and the green band describes the relevant 1$\sigma$ uncertainty. 
This result is an important test of the validity of our initial assumption that 
a MAGN with a radio core emission also emits photons in the $\gamma$-ray energy band, via likely SSC and EC processes.
It is remarkable that the band is a good fit to {\it Fermi}-LAT data.

\begin{figure}
 \includegraphics[width=\columnwidth]{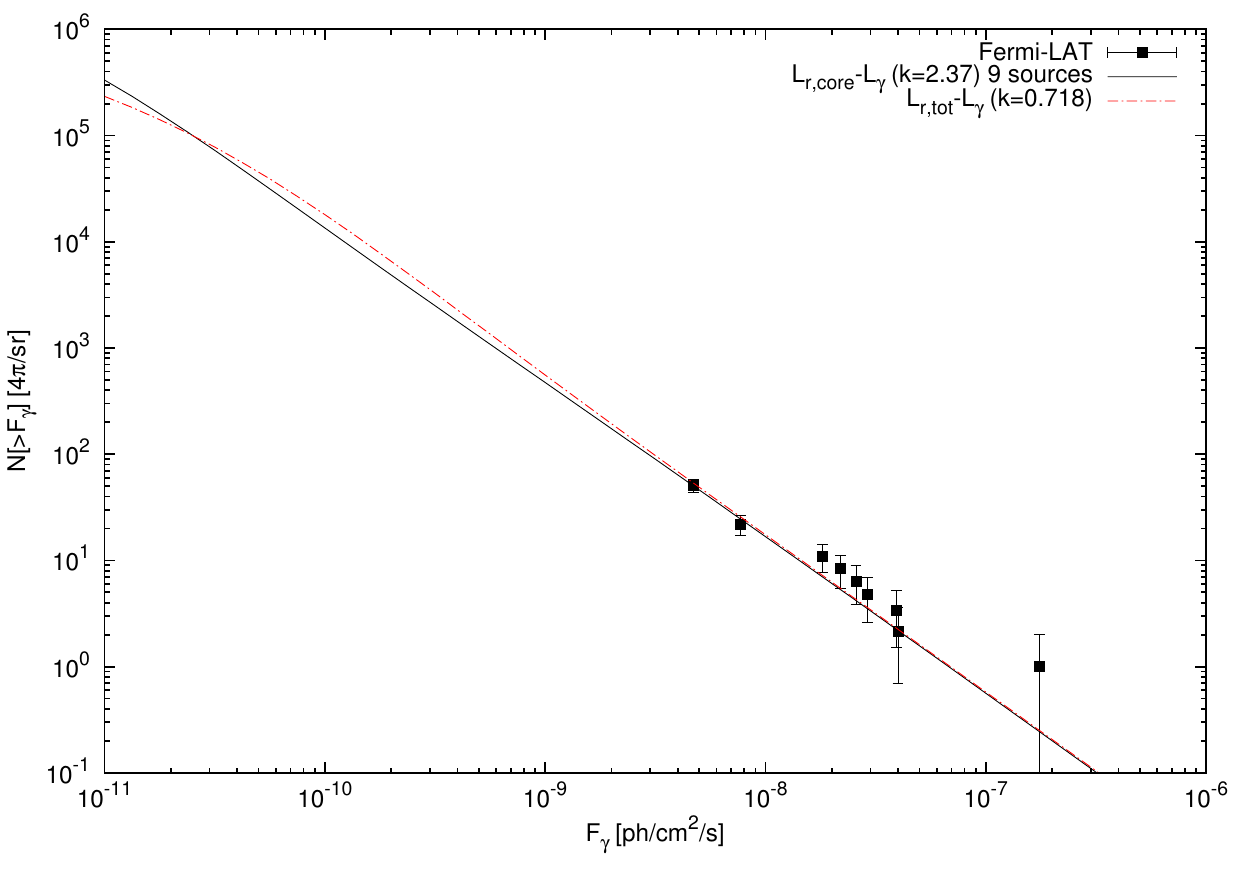}
\caption{ Source count distribution as a function of the integrated $\gamma$-ray flux for 9 RGs. 
{\it Fermi}-LAT data are represented by black squares with 1$\sigma$ error bars. 
The black solid line corresponds to the source count distribution predicted
with the best fit configuration  for the $L_{r, {\rm core}}-L_\gamma$ correlation in Eq. \ref{eq:correlation9} and $N(>F_{\gamma})$. 
The red dot-dashed curve has been obtained with the  $L_{r, {\rm tot}}-L_\gamma$ correlation according to Eq.\ref{eq:correlationtot9}.
\label{fig:ncount9} }
\end{figure}

Given the uncertain classification of some of the sources, as explained in Sec.\ref{sec:correlation}, 
we also provide the source count distribution for 
the 9 sources with firm FRI or FRII classification, Fig.\ref{fig:ncount9}. 
We show the experimental and the theoretical source count distribution predicted when the three galaxies 3C 380, 3C 207 and PKS 0625$-$35 
are excluded from the analysis. The black solid  line is the same as in Fig. \ref{fig:ncount}, but obtained with 9 data points and employing Eq. \ref{eq:correlation9}
for the $L_{r, {\rm core}}-L_\gamma$ luminosity correlation function. 
The result is compatible with data, with $\chi^{2}$ =4.65 and the normalization for the source number distribution $k$=2.37.  
The red dot-dashed curve is the same as in Fig. \ref{fig:ncount}, but obtained from the total RLF and  Eq.\ref{eq:correlationtot9}, 
and minimized with respect to the 9 data points. 

\section{The diffuse $\gamma$-ray emission from MAGN}
\label{sec:flux}
The diffuse $\gamma$-ray flux due to the whole population of MAGN may be estimated as follows:
    \begin{eqnarray} 
    \label{egrbdef}
        &&\frac{d^2F(\epsilon)}{d\epsilon d\Omega} =  \int^{\Gamma_{max}}_{\Gamma_{min}}d\Gamma \frac{dN}{d\Gamma}
	  \int^{z_{max}}_0 \frac{d^2V}{dz d\Omega} dz \int^{L_{\gamma,max}}_{L_{\gamma,min}} \frac{d F_{\gamma}}{d\epsilon} \;  \\
	   &\cdot& \;\frac{dL_{\gamma}}{L_{\gamma} \ln(10)} \; \rho_{\gamma}(L_{\gamma},z) (1-\omega({F_{\gamma}(L_{\gamma},z)}))\;\exp{(-\tau_{\gamma,\gamma}(\epsilon,z))}.  \nonumber
    \end{eqnarray}
The minimum $\gamma$-ray luminosity value is set to $10^{41}$ erg s$^{-1}$,  the maximum at $10^{50}$ erg s$^{-1}$.
The term $\omega(F_{\gamma} (L_{\gamma} , z))$ is the detection efficiency of {\it Fermi}-LAT at the photon flux $F_{\gamma}$,
which corresponds to the flux from a source with a $\gamma$-ray luminosity $L_{\gamma}$ at redshift z.
$dN/d\Gamma$ is the  photon spectral index distribution (see Eq.\ref{Ncount}). 
$d F_{\gamma}/d\epsilon$ is the intrinsic photon flux at energy $\epsilon$, for a MAGN with $\gamma$-ray luminosity $L_{\gamma}$ \citep{venters2009,yan2012}:
    \begin{equation} \label{Fdef}
         \frac{d F_{\gamma}}{d\epsilon} = \frac{(1+z)^{2-\Gamma}}{4\pi {d_L(z)}^2} 
	  \frac{(2-\Gamma)}{\left[\left( \frac{\epsilon_2}{\epsilon_1} \right)^{2-\Gamma}-1\right]}
	  \left( \frac{\epsilon}{\epsilon_1} \right)^{-\Gamma} \frac{L_{\gamma}}{\epsilon^2_1}.
    \end{equation}
High-energy $\gamma$ rays ($\epsilon > 20$ GeV) propagating in the Universe are absorbed by the interaction with the extragalactic background light (EBL), 
cosmic optical radiation and infrared background \citep{1966PhRvL..16..252G,1966PhRvL..16..479J,1992ApJ...390L..49S,
salamon1998,stecker2006,mazin2007,razzaque2009,gilmore2009,finke2010,Ackermann:2012sza,2013A&A...550A...4H},
with an  optical depth  $\tau_{\gamma,\gamma} (\epsilon, z)$. 
In this study we adopt the attenuation model of  \cite{finke2010}. The $\gamma$-ray absorption creates electron-positron pairs, 
which can scatter off the CMB photons through inverse Compton (IC) yielding a 
secondary cascade emission at lower  $\gamma$-ray energies.
We include the cascade emission from high-energy $\gamma$-rays following Refs. \citep{2012PhRvD..86b3003I,2008AA...479...41K} and 
 accounting for the first generation of electrons produced from the interaction of $\gamma$-rays with the EBL.
(In the considered energy range the correction for the second generation of electrons is negligible). 
We assume a maximum $\gamma$-ray energy of 10 TeV as this is the indicative largest energy sampled by current generation TeV telescopes \citep{tevcat,2013AA...554A..75S}.
At these energies, the interaction with the CMB photons is well described by Thomson scattering.
\newline
Within these hypotheses,  the cascade emission is computed according to  Eq.~\ref{egrbdef},
 where the intrinsic photon flux $d F_{\gamma}/d\epsilon$ is replaced by 
\begin{equation} 
\label{cascone}
         \frac{d F^{\rm{casc}}_{\gamma}}{d\epsilon}(\epsilon,z) = \frac{(1+z)}{4\pi {d_L(z)}^2} \int_{\gamma_{e,min}}^{\gamma_{e,max}} \frac{dN_{\gamma_e\, \epsilon}}{dt d\epsilon} 	
	 \frac{dN_e}{d\gamma_e} t_{IC}(z) d\gamma_e,
    \end{equation}
where $t_{IC}(z)$ is the energy-loss time of an electron with a Lorentz factor $\gamma_e$. The term 
$dN_{\gamma_e\, \epsilon}/dt d\epsilon$ is the IC scattered photon spectrum per unit time:
\begin{equation} \label{casctwo}
         \frac{dN_{\gamma_e\, \epsilon}}{dt d\epsilon}  = \frac{3 \sigma_T c}{4\gamma^2_e} \int_{0}^{1} \frac{dx}{x}\frac{dn_{CMB}}{dx} (x(\xi,\gamma_e),z) f(x),
    \end{equation}
where $\sigma_T$ is the Thomson scattering scross section,   $f(x) = 2x\ln{x}+x+1-2x^2$ $(0 < x < 1)$ and $x = \epsilon_{\gamma,i}/4\gamma^2_e \xi$. 
Here, $\epsilon_{\gamma,i} = 2\gamma_e m_e c^2$ is the energy of intrinsic photons and $dn_{CMB}/d\xi$ is the CMB photon density with energy $\xi$. 
The integration  in Eq. \ref{cascone} runs from $\gamma_{e,min}= \max[(E_{\gamma}/\epsilon)^{1/2}/2,100 \, {\rm{MeV}}/2 m_e c^2]$ to 
$\gamma_{e,max} = E_{max}/2m_e c^2$ . 
The electron spectrum $dN_e/d\gamma_e$ is given by:
\begin{equation} \label{cascthree}
         \frac{dN_e}{d\gamma_e} = \frac{d_L(z)^2}{(1+z)} \frac{d\epsilon_{\gamma,i}}{d\gamma_e} \frac{d F_{\gamma}}{d\epsilon} \left( 1 - \exp{(-\tau_{\gamma,\gamma})}\right),
    \end{equation}
where $d F_{\gamma}/d\epsilon$ is given by Eq.~\ref{Fdef}.

 \begin{figure*}
 \begin{centering}
 \includegraphics[]{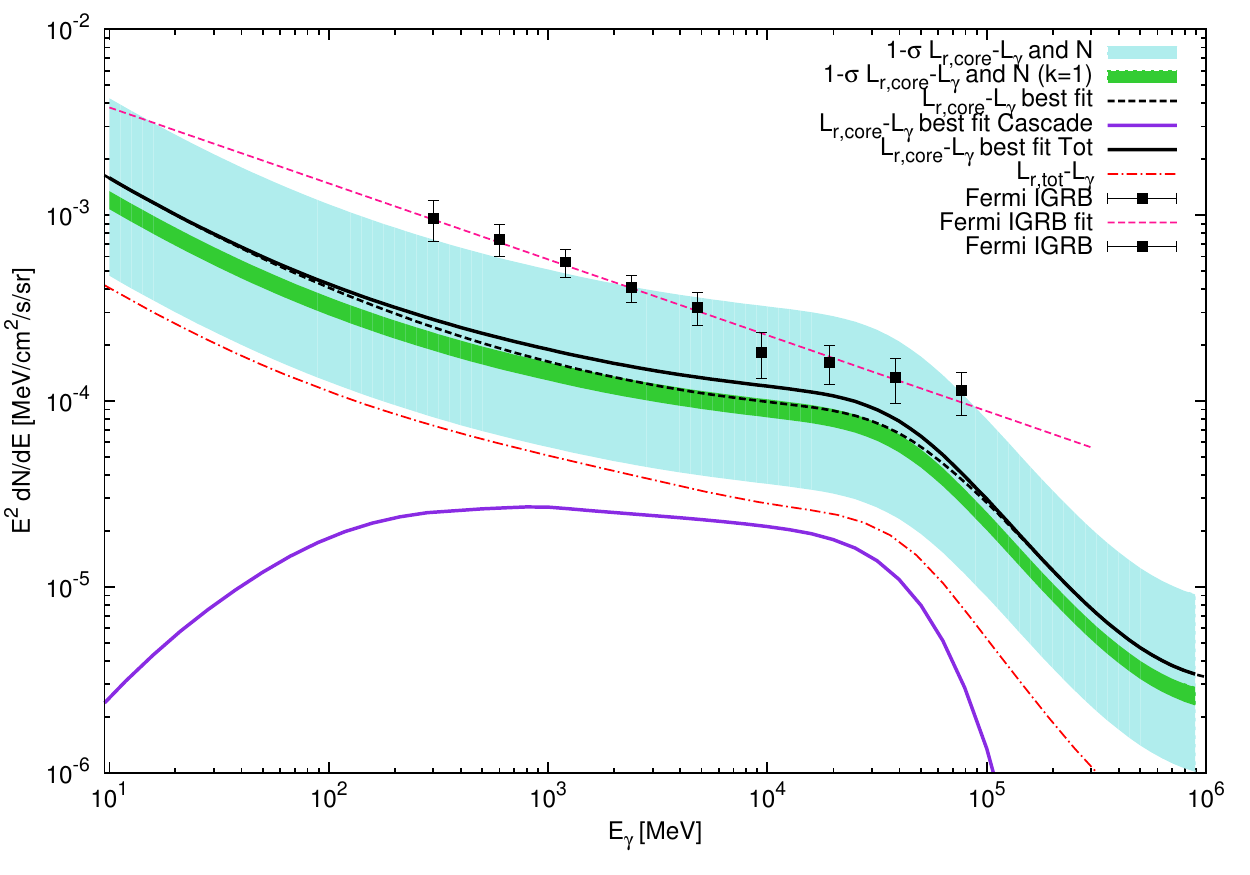}
\caption{Diffuse  $\gamma$-ray  flux due to the MAGN as a function of $\gamma$-ray energy. 
The black squares correspond to the IGRB measured by {\it Fermi}-LAT \citep{IDGRB} and best-fitted by  the magenta dashed curve. 
The cyan shaded area  derives from the 1$\sigma$ uncertainty band on the $L_{r, {\rm core}}-L_\gamma$ correlation
and on the $k$ parameter in the source count distribution.
The solid  black line is obtained from the best fit on the $L_{r, {\rm core}}-L_\gamma$ correlation, 
while the dashed black line illustrates the relevant flux without the contribution from the cascade emission, 
depicted by the violet solid line.
The green band corresponds to $k$=1 at 1$\sigma$ C.L. (see discussion on Fig.\ref{fig:ncount}).
The red  dot-dashed curve shows the diffuse flux obtained when assuming a  $L_{r, {\rm tot}}-L_\gamma$ correlation.
\label{fig:flux}}
\end{centering}
\end{figure*}

\noindent
Fig.\ref{fig:flux} shows the diffuse $\gamma$-ray  flux due to the MAGN population as a function of $\gamma$-ray energy, 
along with the {\it Fermi}-LAT data for the IGRB \citep{IDGRB}. 
The cyan shaded area  derives from the 1$\sigma$ uncertainty band on the $L_{r, {\rm core}}-L_\gamma$ correlation
and on the $k$ parameter in the source count distribution  (see description of the cyan shaded area  in Fig. \ref{fig:ncount}).
The upper edge of the uncertainty band skims the IGRB data points, while the 
lower limit is almost an order of magnitude below the data. The band
itself is nearly a factor of ten wide. 
The  flux integrated above 100 MeV is $5.69\cdot10^{-7}$ cm$^{-2}$s$^{-1}$sr$^{-1}$ for the lower bound of the 
uncertainty band, and $4.91\cdot10^{-6}$ cm$^{-2}$s$^{-1}$sr$^{-1}$ for the upper one. These values compare with 
$1.03\cdot10^{-5}$ cm$^{-2}$s$^{-1}$sr$^{-1}$ derived from the experimental data \citep{IDGRB}.
 The green band has been obtained by fixing $k=1$ as described in Sect. \ref{sec:Ncount}
on Fig. \ref{fig:ncount}.  It corresponds to the  case in which all the MAGN with  
a radio-loud central region emit in $\gamma$ rays as well, and with a phenomenological model that fits nicely all the experimental constraints. 
\\
The flux calculated for the best fit coefficients of the $L_{r, {\rm core}}-L_\gamma$ correlation and  $k$=3.05
(see description of the  black dashed line in Fig. \ref{fig:ncount}) is displayed in Fig. \ref{fig:flux}. 
The corresponding cascade emission is illustrated as a violet curve. It  shows a flat behavior (with respect to the $E^2$ normalization 
adopted in the figure) between about 200 MeV and 30 GeV, while it drops sharply at higher energies. At 100 MeV/1 GeV/10 GeV 
the cascade emission is about 4.5\%/16\%/21\% of the non-absorbed flux. 
As a consistency check, we estimated the energy flux of the photons absorbed by interaction with the EBL and compared it to that from the
cascade emission, as well as the total energy flux from the photons
arriving at the LAT which are not absorbed by EBL (dashed black line in Fig. \ref{fig:flux}).
This latter is obtained by integrating Eq. 23 multiplied by the energy, between $100$ MeV and $10$ TeV, and has a value of $2.35\cdot10^{-3}$ MeV cm$^{-2}$sr$^{-1}$ s$^{-1}$ .
By simply replacing the EBL attenuation term
$\exp{(-\tau_{\gamma,\gamma}(\epsilon,z))}$ with its complement $(1 - \exp{(-\tau_{\gamma,\gamma}(\epsilon,z))})$, and performing the same
integral, we computed the energy flux of those photons that get absorbed by the EBL and can be reprocessed through the cascade,
obtaining a value of $2.96\cdot10^{-4}$ MeV cm$^{-2}$sr$^{-1}$s$^{-1}$. 
This can be considered as an upper limit to the cascade emission, and is in fact slightly higher than its actual flux of $1.93\cdot10^{-4}$ MeV cm$^{-2}$sr$^{-1}$ s$^{-1}$, which anyway represents only 8\% of the total MAGN flux.


Our predictions are for a MAGN population whose $\gamma$-ray emission is assumed to originate from the central region of the 
active galaxy, and modeled from the core RLF. 
The dot-dashed red line represents the  flux derived when the $\gamma$-ray luminosity is correlated with the total radio luminosity
according to Eq. \ref{eq:correlationtot}, and total RLF \citep{willott}  (see description of the red dot-dashed line in Fig. \ref{fig:ncount}).
The effect of EBL absorption is clear from the softening of the flux above 50 GeV.
The deviation from a pure power-law shape below $\sim 30$ GeV is due to integration over the photon index distribution. 
We note that the contribution of unresolved blazars  \citep{2010ApJ...720..435A} has a very similar slope but is lower than 
the one obtained for  MAGN in this paper. The two uncertainty bands nearly touch each other. 
\cite{inoue} reported that MAGN can contribute to the IGRB at the level of 10-63 \%, which is a
range compatible with our result.
\\
The flux displayed in Fig. \ref{fig:flux}  results from an integration up to a maximum luminosity of $10^{50}$ erg s$^{-1}$.
The result does not depend on the maximal luminosity of integration, confirming that
the photons come from very numerous and very faint sources. A confirmation of the negligible contribution of bright sources
to the overall flux is that the flux at 1 GeV (multiplied by $E^2$) for the 15 galaxies of Table \ref{tab:galaxies}  
is $3.5\cdot 10^{-6}$ MeV cm$^{-2}$ s$^{-1}$ sr$^{-1}$, 
more than two orders of magnitude  below our estimated diffuse flux. We finally observe that shifting  the lower luminosity from
10$^{41}$ erg s$^{-1}$ down to 10$^{38}$ erg s$^{-1}$ would lead to a 15\% greater isotropic intensity. 
\\
Our predictions may be compared to the results reported by \cite{inoue}. In that paper, the flux from unresolved MAGN
has been obtained for a single  model, which is contained in our uncertainty band and shows a different shape 
witfh respect to our representative cases (solid and dashed lines in Fig. \ref{fig:flux}).
Possible differences between the two procedures are probably due to the fact that \cite{inoue} works within a smaller and different RG sample.
We both establish a $L_{r, {\rm core}}\; - \; L_\gamma$ correlation.  However, we convert the total RLF by \cite{willott} to core RLF, while 
\cite{inoue} does not make  this transformation.
Data on the source number distribution are different, in particular the {\it Fermi}-LAT data points at the lowest fluxes. 
A final possible difference might reside in a  different angular conversion factor in the RLF coefficients 
in \cite{willott}.


\section{Conclusions}
\label{sec:conclusions}
We have calculated the diffuse $\gamma$-ray emission from the
population of  MAGN at all redshifts.
\\
We first established the existence (at 95\,\% C.L.)  of a correlation between the radio core
($L_{r, {\rm core}}$) and the $\gamma$-ray ($L_\gamma$) luminosities 
of the MAGN detected by the {\it Fermi}-LAT. 
This correlation is substantially linear in the log plane, the radio luminosity being two orders of magnitude
 lower than the $\gamma$-ray luminosity. 
 Extensive tests showed that this correlation is not likely to
be a spurious effect due to the  source distance.
We also calculated the upper limits on the $\gamma$-ray emission from
33 radio-loud MAGN undetected by {\it Fermi}-LAT, and showed that these are compatible 
with the core radio $-$ $\gamma$-ray luminosity correlation within 1$\sigma$ errors. 
\\
We then used this correlation to infer a $\gamma$-ray luminosity function from a well established radio
luminosity function, and further tested the former against the source count distribution measured by the {\it Fermi}-LAT. 
We correctly predicted the number of detected $\gamma$-ray sources, with
values of the normalization factor $k$ between the population of MAGN
emitting in radio and $\gamma$ rays that are close to one.
Even when constraining $k=1$, our $\gamma$-ray luminosity function
matched the {\it Fermi}-LAT source count distribution, nicely
confirming the robustness and simplicity of the luminosity correlation we derived.
\\
Using our $\gamma$-ray luminosity function, and after taking into account $\gamma$-ray absorption from a
model of EBL, we predicted the diffuse $\gamma$-ray flux due to  MAGN between 10 MeV and 1 TeV.
We found an intensity of about $2\cdot
10^{-4}$ MeV cm$^{-2}$s$^{-1}$sr$^{-1}$ at 1 GeV, embedded in a 
uncertainty band of nearly a factor of ten.
At all {\it Fermi}-LAT energies, the best fit MAGN contribution is
20\%-30\% of the measured IGRB flux.
The lower edge of the uncertainty band is about one order of magnitude smaller than the IGRB data
while the upper edge skims the data below a few GeV and 
slightly over-estimates them from a few GeV to around 50 GeV. 
Our uncertainty band includes the results found by \cite{inoue}, 
based on a correlation between $\gamma$-rays and the total radio luminosity. 
At higher energies, the flux softens because of the EBL absorption. 
The intensity from MAGN integrated above 100 MeV is $9.83\cdot 10^{-7}/2.61\cdot 10^{-6}/8.56\cdot 10^{-6}$ photons cm$^{-2}$ s$^{-1}$  sr$^{-1}$, 
when considering the lower/best fit/upper curve of the band reported in Fig. \ref{fig:flux}. These numbers represent 
 9.5\%/25\%/83\% of the IGRB, respectively. The analogous calculation for the two blazar populations of BL Lacs and 
FSRQs gives $7.83^{+1.09}_{-2.34}\cdot 10^{-7}$ photons cm$^{-2}$ s$^{-1}$  sr$^{-1}$ (about 8\% of the IGRB) 
for the former \cite{2010ApJ...720..435A}
and $9.66^{+1.67}_{-1.09} 10^{-7}$ photons cm$^{-2}$ s$^{-1}$  sr$^{-1}$ (about 10\% of the IGRB) for the latter \cite{2012ApJ...751..108A}.
The integrated flux for star-forming galaxies \cite{2012ApJ...755..164A} is 
instead $8.19^{+7.31}_{-3.89} \cdot 10^{-7}$ photons cm$^{-2}$ s$^{-1}$  sr$^{-1}$, which contributes 
 4.1\% (14.8\%) of the IGRB at minimum (maximum), and about 8\% at its best fit value. 
\\
In conclusion,we have calculated the diffuse $\gamma$-ray flux from unresolved MAGN. 
The main original results of our analysis include i) the derivation of 
a $\gamma$-ray - radio core luminosity correlation for the MAGN observed by {\it Fermi}-LAT; 
ii) the test of this correlation against upper limits from tens of radio loud MAGN undetected in $\gamma$-rays; 
iii) tests of the correlation in order to verify that the radio core - $\gamma$-ray luminosity correlation for
MAGN is not spurious; 
iv) the calculation of the $\gamma$-ray luminosity function from the $core$ radio one; v) evaluation of the 
uncertainties affecting $\gamma$-ray flux predicted from the unresolved MAGN population.
We have found that the cosmological population of faint and numerous MAGN gives a sizable
diffuse extragalactic flux which, when added to the contribution from other  sources ($i.e.$ blazars
\citep{2010ApJ...720..435A}, star-forming galaxies \citep{2012ApJ...755..164A}, 
millisecond pulsars \citep{FaucherGiguere:2009df,SiegalGaskins:2010mp}, cascade from 
ultra-high energy cosmic rays \citep{2010APh....34..106A}, Radio-quiet AGNs \citep{2009ApJ...702..523I}, 
large scale structures \citep{2003APh....19..679G}, strong galactic foreground \citep{2004JCAP...04..006K},
 cosmic-ray interaction in the Solar System \citep{2009ApJ...692L..54M}), could entirely explain the observed IGRB. 
This scenario would leave very little room for more exotic sources, such as dark matter in the halo of our Galaxy \citep{Calore:2013yia}.


\appendix
\label{appendix}
\section{Estimation of the  2FGL efficiency}
Out of the 12 MAGN considered in our analysis, 
8 galaxies (3C78, 3C274, Cen A, NGC 6251, PKS 0625-35, 3C111, 3C207, 3C380)
are found in the 1FGL, 
8 (3C274, Cen A, NGC 6251, Cen B, Fornax A, PKS 0625-35,  3C207, 3C380) are in the 2FGL, 
while 3C120 is listed in \cite{misaligned} and  Pictor A has been revealed in \cite{pictora}.  
The efficiency employed in our analysis is taken from \citep{2010ApJ...720..435A}, which refers to the 1FGL blazar catalog.
Since the source detection efficiency was not published for the 2FGL, we have assumed in this paper that the same 1FGL efficiency holds for all the MAGN 
in Table \ref{tab:galaxies}. 
In this Section we infer an efficiency for the 2FGL catalog and check if the logN-logS associated to the 
2FGL sources is consistent with the results discussed in Sect. \ref{sec:Ncount}. 
\\
We start from the blazar logN-logS count distribution established in \cite{2010ApJ...720..435A}.
As a first step, we search all the 2FGL blazars with TS$>$25 (in accordance with the MAGN TS)
 and $|b|>10^\circ$ (in order to exclude the contamination from the galactic plane). 
For all the selected sources, we compute the flux  $F_{\gamma}$  from 100 MeV up to 100 GeV according to Eqs. \ref{Sg},\ref{dNde}. 
We have considered a  flux range from $10^{-9}$ ph/cm$^2$/s to $10^{-5}$ ph/cm$^2$/s divided in N bins. 
The efficiency $\omega(F^k_{\gamma})$ at a flux $F^k_{\gamma}\in (F^{k,min}_{\gamma},F^{k,max}_{\gamma})$ ($k=\{1,.....,N\}$)
may be estimated as:
\begin{equation} 
\label{efficiency}
         \omega(F^k_{\gamma})=(1+\eta) \frac{N_{\rm blazars}^k}
         {\Delta\Omega \int^{F^{k,max}_{\gamma}}_{F^{k,min}_{\gamma}} \frac{dN}{dF_{\gamma}}dF_{\gamma}}
    \end{equation}
where $\Delta\Omega$ is the solid angle associated to $|b|>10^{\circ}$ and $N_{\rm blazars}^k$ is the number of selected blazars with flux
 $F_{\gamma}\in (F^{k,min}_{\gamma},F^{k,max}_{\gamma})$. The integrand $dN/dF_{\gamma}$ is the flux distribution of blazars, 
 and the term $\int^{F^{k,max}_{\gamma}}_{F^{k,min}_{\gamma}} \frac{dN}{dF_{\gamma}} \,dF_{\gamma}$ in Eq. \ref{efficiency} represents 
the expected number of blazars. The incompleteness of the 2FGL catalog $\eta$ 
is given by the ratio of unassociated sources to the total number of sources. 
In the 2FGL (for TS$>25$ and $|b|>10^\circ$)  there are 1042 sources out of which 169 are unassociated, giving $\eta\approx0.16$.
The flux distribution $dN/dF_{\gamma}$ of the 2FGL being unknown, we assume it to be the one from the 1FGL
taken  from \cite{2010ApJ...720..435A}  (for TS$>50$ and $|b|>20^\circ$). 
\begin{figure}[t]
\begin{minipage}[b]{0.45\linewidth}
\centering
 \includegraphics[scale=0.7]{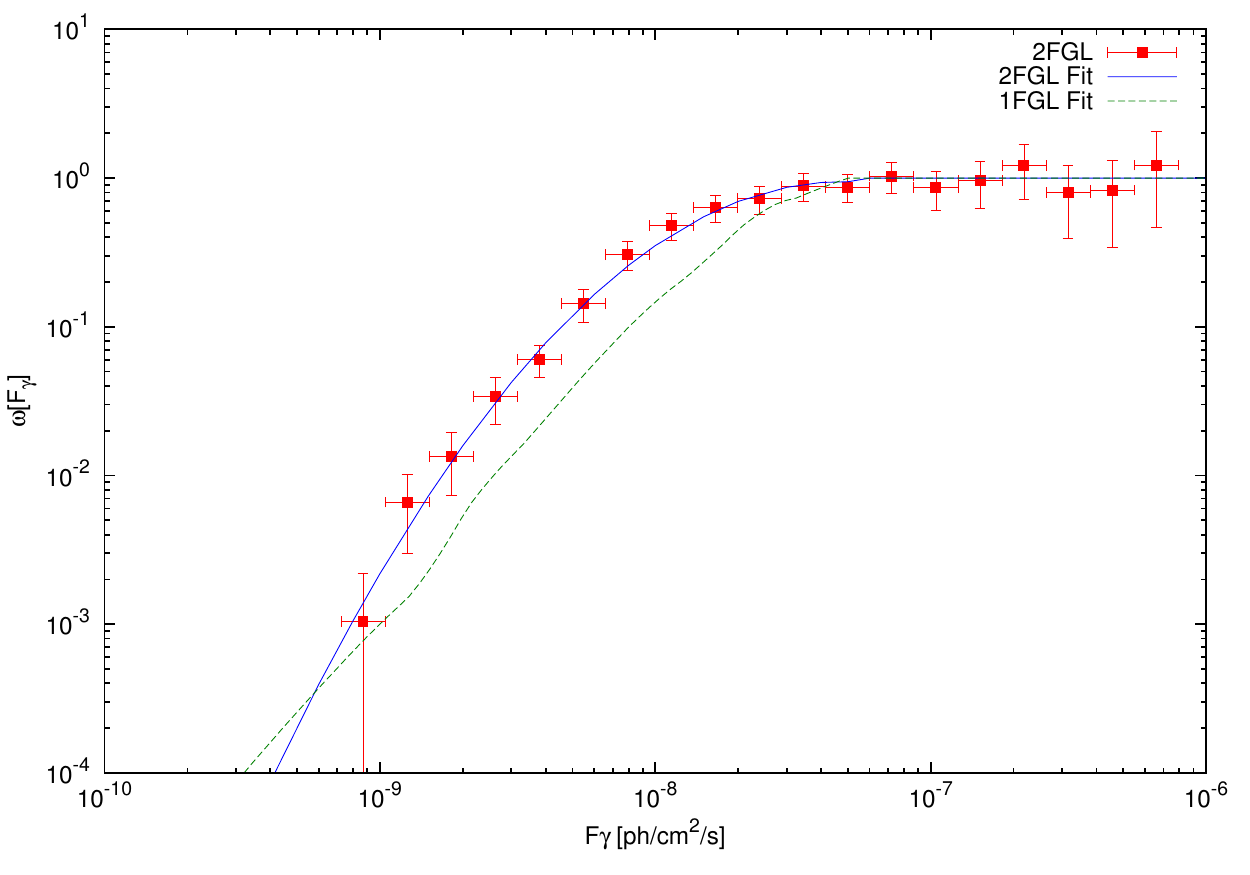}
\caption{Evaluated efficiency for the 2FGL (red points fitted by the blue curve) along with the 1FGL efficiency 
\citep{2010ApJ...720..435A} (green  dashed curve), as a function of the integrated $\gamma$-ray flux $F_{\gamma}$. }
\label{fig:efficiency} 
\end{minipage}
\hspace{0.5cm}
\begin{minipage}[b]{0.45\linewidth}
\centering
 \includegraphics[scale=0.7]{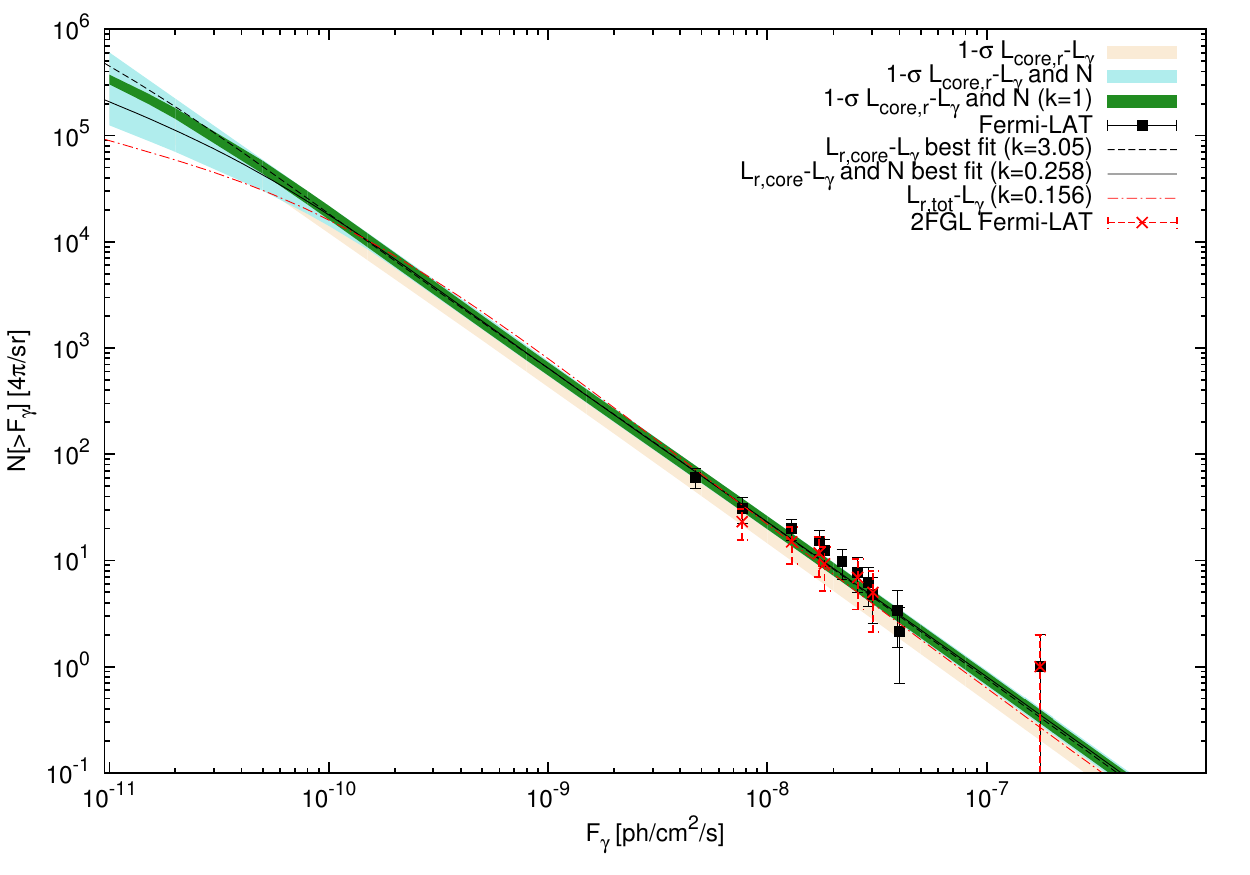}
\caption{Same as in Fig. \ref{fig:ncount}, with the addition of the red points 
indicating the number count evaluated with the efficiency estimated for the 2GL (as in Fig. \ref{fig:efficiency}), 
 for the 8 MAGN detected in the 2FGL catalog.}
\label{ncount_2cat}
\end{minipage}
\end{figure}
\\
In  Fig. \ref{fig:efficiency} we show the estimated  efficiency found with the method described here.
The error bars represent the uncertainties on $dN/dF_{\gamma}$ and the Poisson errors associated to the observed number of blazars
 $N_{\rm blazars}^k$  counted in each flux bin. We also overlap the 1FGL efficiency  \citep{2010ApJ...720..435A}. 
 The two efficiencies are similar at high fluxes $F_{\gamma} > 4\cdot 10^{-8}$ ph/cm$^2$/s, while 
 at lower values the 2FGL efficiency is shifted to the left side. 
This is due to the selection criteria used for deriving the efficiency, which are looser for 2FGL 
  (TS$>25$ and $|b|>10^\circ$) with respect to the 1FGL  (TS$>50$ and $|b|>20^\circ$).
  \\
Finally, in Fig. \ref{ncount_2cat} we  display the logN-logS  as in Fig.  \ref{fig:ncount} (for 12 MAGN) to which we add 
the (red) points corresponding to the number count computed for  the 8 MAGN in the 2FGL catalog and 
assuming the detection efficiency estimated for the 2FGL as in Fig.  \ref{fig:efficiency}. 
We can conclude that the source number count for the 2FGL sample and with newly estimated 
efficiency is compatible with the results obtained for the whole MAGN treated with the 1FGL efficiency. 
Additionally, a slightly different  normalization can be safely compensated by the free normalization parameter $k$ 
(see discussion in Sect. \ref{sec:Ncount}) and will not change  the flux prediction  
derived in Sect. \ref{sec:flux}.


\begin{acknowledgements}
M.D.M. and F.D. warmly acknowledge S. Massaglia for invaluable insights in the preliminary stage of this paper
and P.D. Serpico for useful comments. 
M.A. acknowledges support from grant NNH09ZDA001N for the study of the origin of the Isotropic Gamma-ray 
Background. F.C. acknowledges support from the German Research Foundation (DFG) through grant BR 3954/1-1.
\\
The {\it Fermi} LAT Collaboration acknowledges generous ongoing sup- port from a number of
agencies and institutes that have supported both the development and the operation
of the LAT as well as scientific data analysis. These include the National
Aeronautics and Space Administration and the Department of Energy in the United
States; the Commissariat  \`a l'Energie Atomique and the Centre National de la
Recherche Scientifique/Institut National de Physique Nucl\'eaire et de Physique des
Particules in France; the Agenzia Spaziale Italiana and the Istituto Nazionale di
Fisica Nucleare in Italy; the Ministry of Education, Culture, Sports, Science and
Technology (MEXT), High Energy Accelerator Re- search Organization (KEK), and Japan
Aerospace Exploration Agency (JAXA) in Japan; and the K. A. Wallenberg Foundation,
the Swedish Research Council, and the Swedish National Space Board in Sweden.
Additional support for science analysis dur- ing the operations phase is gratefully
acknowledged from the Istituto Nazionale di Astrofisica in Italy and the Centre
National d'Etudes Spatiales in France. 
\end{acknowledgements}

\bibliography{paper}

\end{document}